\documentclass[twocolumn, tighten]{aastex7}

\begin{document}

\title{A Hybrid Framework for Kilonova Anomaly Detection using Single-Epoch SEDs from the 7-Dimensional Telescope}

\author[orcid=0000-0002-6639-6533,gname='Gregory', sname='Paek']{Gregory S. H. Paek}  
\affiliation{Institute for Astronomy, University of Hawaii, 2680 Woodlawn Drive, Honolulu, HI 96822, USA}
\affiliation{SNU Astronomy Research Center, Astronomy Program, Department of Physics \& Astronomy, Seoul National University, 1 Gwanak-ro, Gwanak-gu, Seoul,  08826, Republic of Korea} 
\email{gregorypaek94@gmail.com} 

\correspondingauthor{Myungshin Im}
\author[orcid=0000-0002-8537-6714,gname=Myungshin, sname='Im']{Myungshin Im} 
\affiliation{SNU Astronomy Research Center, Astronomy Program, Department of Physics \& Astronomy, Seoul National University, 1 Gwanak-ro, Gwanak-gu, Seoul,  08826, Republic of Korea} 
\email[show]{myungshin.im@gmail.com}

\author[orcid=0000-0002-3118-8275,gname='Seo-Won', sname='Chang']{Seo-Won Chang}
\affiliation{SNU Astronomy Research Center, Astronomy Program, Department of Physics \& Astronomy, Seoul National University, 1 Gwanak-ro, Gwanak-gu, Seoul,  08826, Republic of Korea}
\email{seowon.chang@snu.ac.kr} 

\author[orcid=0000-0003-4422-6426,gname='Hyeonho', sname='Choi']{Hyeonho Choi}
\affiliation{SNU Astronomy Research Center, Astronomy Program, Department of Physics \& Astronomy, Seoul National University, 1 Gwanak-ro, Gwanak-gu, Seoul,  08826, Republic of Korea}
\email{hhchoi1022@gmail.com}

\author[orcid=0000-0002-1418-3309,gname='Jihoon', sname='Kim']{Ji Hoon Kim}
\affiliation{SNU Astronomy Research Center, Astronomy Program, Department of Physics \& Astronomy, Seoul National University, 1 Gwanak-ro, Gwanak-gu, Seoul,  08826, Republic of Korea}
\email{jhkim.astrosnu@gmail.com}

\begin{abstract}

We develop a hybrid framework to identify kilonovae (KNe), using single-epoch, medium-band spectral energy distributions from the 7-Dimensional Telescope (7DT). 
The framework integrates an unsupervised anomaly classifier (\texttt{Isolation Forest}) to flag unusual events with a supervised multi-class classifier (\texttt{XGBoost}) that characterizes eight common transient types. 
Trained on realistically simulated 7DT photometry accounting for per-filter sensitivity, the classifier achieves macro $F_{1}\sim0.80$ ($\sim0.82$) with 20 (40) filters across eight classes, Type~Ia/Ibc/II SNe, SLSNe, TDEs, AGN, stellar variables, and asteroids. 
Without direct training, the anomaly detector recovers $>$90\% of simulated and observed optically detectable KNe (AT~2017gfo) with a low contamination fraction, with a caveat of limitations of the training sample such as limited redshift range of SNe ($z < 0.15$), and a relatively small number of early non-KNe spectra.
A SHAP-based feature analysis reveals that only $\sim$40–50\% of the most informative filters are sufficient to retain near-baseline performance, while red-end filters contribute little. 
Combining the top-ranked half of the 40 7DT filters with a single LSST band reproduces the full-model accuracy within 1–2\%, suggesting practical follow-up strategies. 
These results demonstrate that 7DT's medium-band system enables rapid, interpretable classifications and reliable anomaly alerts from single-epoch data---promising for gravitational-wave follow-up, Rubin alert stream filtering, and serendipitous transient discovery in the 7DT survey.

\end{abstract}



\section{Introduction}\label{sec:intro}

We are entering an unprecedented era of time-domain astronomy, defined by the upcoming \textit{Legacy Survey of Space and Time} (LSST; \citealt{2019ApJ...873..111I}) conducted by the Vera C. Rubin Observatory. LSST will produce up to $\sim10^7$ alerts per night, unveiling vast populations of both common and rare transients. Identifying scientifically valuable rare events rapidly and reliably within this data torrent represents one of the most pressing challenges in modern astronomy. This new landscape demands scalable, efficient, and robust classification frameworks capable of isolating rare sources of interest from overwhelming background populations.

Kilonovae (KNe), the electromagnetic (EM) counterparts to gravitational-wave (GW) sources such as binary neutron star (BNS) mergers, exemplify this challenge. The detection of GW170817 and its optical counterpart AT~2017gfo \citep{2017ApJ...848L..12A,2017ApJ...848L..14G,2017ApJ...848L..15S,2017Sci...358.1556C,2017Natur.551...71T} established KNe as a distinct class of fast-evolving transients powered by the radioactive decay of heavy $r$-process nuclei \citep{2010MNRAS.406.2650M,2012ApJ...746...48M,2017Natur.551...71T,2017Natur.551...80K,2024arXiv240910651K}. However, searches for KNe remain difficult due to the large GW localization areas and their intrinsically faint, rapidly fading nature. Compounding the problem, the vast majority of candidates identified in third and fourth runs of LIGO/Virgo/KAGRA Collaboration were non-KN contaminants-supernovae (SNe), Active Galactic Nuclei (AGN), Cataclysmic Variables (CVs), flare stars, and asteroids \citep[e.g.,][]{2021MNRAS.507.1296O,2022ApJ...927...50R,2024PASP..136k4201A,2020MNRAS.499.3868T}. These same populations are expected to dominate LSST alert streams, making KN searches a natural testbed for classification strategies in the Rubin era.

Two principal approaches have been developed for classifying optical transients.
Spectroscopy classifies transients by identifying the overall shape of SEDs, diagnostic line profiles, ionization states, and ejecta velocities.
On the other side, multi-epoch photometric classification relies on the temporal evolution of broad-band or multi-color photometry.

While spectroscopy provides the most reliable method, it is often impractical for faint or rapidly evolving sources due to the required signal-to-noise (SNR) ratio and telescope time. 
In practice, spectroscopic classification is commonly implemented via template matching (\texttt{SNID}; \citealt{2007ApJ...666.1024B}) or machine-learning models trained on spectra (\texttt{DASH}; \citealt{2019ApJ...885...85M}, \texttt{ABC-SN}; \citealt{2025arXiv250722106F}). 
\texttt{SNID} remains the baseline tool for SN classification, relying on template matching of high-resolution ($R\!>\!300$) spectra, though its performance has not been formally quantified. 
\texttt{DASH} introduced a deep convolutional neural network trained on $R\!\approx\!700$ SN spectra and demonstrated comparable or better performance than template-based methods, with $F_{1}$ scores ranging from 0.40--0.82 depending on dataset quality, and classification speeds about 100 times faster than \texttt{SNID}. 
The more recent \texttt{ABC-SN} framework applies a Transformer with attention mechanisms, achieving a macro $F_{1}\!\approx\!0.82$ on 10 SN subtypes at $R\!\approx\!100$, and maintaining stable performance even at $R\!\approx\!30$.
In parallel, \citet{2025MNRAS.543..247M} compared these classifiers under realistic 4MOST-like conditions ($R\!\approx\!50$) and found $>\!80$\% accuracy, with Type~\,Ia~\,SNe reaching 99.9\% purity and 70\% completeness when combining \texttt{DASH} and \texttt{NGSF}. 
By contrast, recent benchmarks on low-resolution SN spectra ($R\!<\!100$; \citealt{2024PASP..136k4501K}) report accuracies typically below $80$\%, underscoring the challenges faced by traditional template matching at low SNR and low spectral resolution.
Training on synthetic spectra from theoretical models \citep{2017Natur.551...80K,2019MNRAS.489.5037B,2021ApJ...918...10W} can mitigate data scarcity but may introduce systematic errors owing to uncertainties in the underlying physics. 
For KNe, the situation is especially restrictive: AT~2017gfo remains the only spectroscopically confirmed example to date, raising risks of overfitting and bias when applying such spectroscopic classifiers.

By contrast, multi-epoch photometric classification is generally more cost-effective than spectroscopic classification. Yet the coarse spectral resolution of broad-band filters ($\Delta\lambda \gtrsim 100$~nm) smooths over diagnostic features, and the need for multi-epoch monitoring can delay classification during the critical early phases. 
Moreover, imposters--such as luminous blue variables (LBVs) and some Type~\,IIb~\,SNe can convincingly mimic KN-like light-curve behavior for several days, posing a serious challenge to photometric-only classification \citep{2025MNRAS.542..541F,2025PASP..137h4105B}. 
AT2025ulz--once a leading candidate for the GW–EM counterpart to the BNS merger S250818k \citep{2025GCN.41437....1L,2025GCN.41440....1L}--illustrates this degeneracy: the time-series photometric data in early phase could not distinguish its Type~IIb shock-cooling phase from a possible KN \citep{2025arXiv251001142G,2025arXiv251023732K,2025arXiv251024620H}.
Taken together, these limitations call for a classification capability that identifies sources from a single-epoch ``snapshot'' SED-avoiding the latency of multi-epoch light curves-and thus motivate low-resolution spectro-photometric approaches.


Recent developments in medium- and narrow-band photometric systems such as \texttt{J-PLUS} \citep{2014arXiv1403.5237B} and \texttt{S-PLUS} \citep{2019MNRAS.489..241M} demonstrate that optimized filter sets can isolate key spectral features and achieve an effective balance between spectral resolution and photometric efficiency. Importantly, however, the main limitation for rapid transient follow-up is infrastructural rather than scientific: their single-telescope, single–optical-path architectures acquire many bands sequentially via filter changes. For fast-evolving sources, the inter-filter overheads and time offsets between exposures can introduce color-evolution biases and prevent the construction of a truly single-epoch SED. Therefore, motivating multi-aperture systems that can obtain medium-band coverage simultaneously.

The 7-Dimensional Telescope (7DT) system was developed with time-domain capability by installing multiple telescopes as a core design consideration \citep{2024SPIE13094E..0XK}. 
Its primary science goal is the rapid identification and characterization of GW-EM counterparts (KNe) \citep{2024cosp...45.1744I}, while also supporting a multi-purpose wide-field survey program-the 7-Dimensional Sky Survey (7DS; \citealt{2024cosp...45.1713I})-that includes all-sky stacking and serendipitous transient discovery. 
It employs a unique architecture of 20 simultaneously operated 0.5-m telescopes at the El Sauce Observatory in Chile, each equipped with a motorized filter wheel housing both medium- and broad-band filters. 
Currently, 20 filters spanning $m400$–$m875$ with 25\,nm spacing are deployed on 16 telescopes, with an eventual goal of 40 filters at 12.5\,nm spacing. 
This configuration enables same-epoch low-resolution SED acquisition across the optical range-critical for capturing fast-evolving transients without relying on multi-epoch observations. 
To achieve its core objective of identifying KNe in the context of GW–EM follow-up, as well as to prepare for the LSST era of massive alert streams, it is essential to develop a 7DT-specific single-epoch SED classification framework optimized for rapid triage of transient candidates. 


Recent studies have investigated the feasibility of robust transient classification using low-resolution spectroscopic data and machine-learning applications. For example, \citet{2025MNRAS.543..247M} showed with simulated \texttt{4MOST}-like blended spectra that SN subtypes can be classified with accuracies exceeding 80\% even at spectral resolutions as low as $R\sim50$. Similarly, the attention-based framework \texttt{ABC-SN} \citep{2025arXiv250722106F}, using spectra from \texttt{SEDMv2}, achieved high performance for SN subtypes at $R\sim30$ by leveraging the substantial Doppler broadening of SN ejecta to preserve key features despite coarse resolution. Notably, both efforts focus on supernova classification and do not probe the broader transient zoo; moreover, \texttt{ABC-SN} explicitly defers a comprehensive treatment of wavelength-dependent SNR requirements and operates over a restricted wavelength range of 4500–7000\,\AA. 

Taken together, the studies suggest that low-resolution spectroscopic SEDs are effective for SN subtype work, but robust application to LSST alert processing and GW–EM follow-up demands realistic, wavelength-dependent sensitivity modeling for diverse transients and a dedicated approach to detecting rare KNe. Moreover, to move beyond SN-focused use cases, a more robust classifier that explicitly accommodates non-SN transients---and addresses classes that have proven difficult for multi-epoch photometry (e.g., LBVs)---is warranted.


In this paper, we present a hybrid classification framework that uses single-epoch 7DT SEDs to classify multiple types of transients and efficiently separate KNe from background populations. 
We train a supervised machine-learning classifier on realistic synthetic photometry that includes wavelength-dependent filter sensitivity and represents eight common transient classes, while applying an unsupervised anomaly detector trained solely on non-KN data to identify KNe as out-of-distribution anomalies. 
We further quantify filter importance for the multi-class classifier to propose optimized filter combinations and to outline observing strategies under realistic operational constraints, and we investigate synergies between 7DT and LSST broad-band data to evaluate classification gains when the two are combined.


\section{Data} \label{sec:data}

\begin{deluxetable}{ll}
\tablecaption{Input Settings for the \texttt{7DT-Simulator}\label{tab:setting}}
\tablehead{
\colhead{\textbf{Parameter}} & \colhead{\textbf{Value}}
}
\startdata
\multicolumn{2}{l}{\textit{Telescope / Optics}} \\
Aperture Diameter       & 50.5 cm \\
Pixel Scale             & 0.51~arcsec~pixel$^{-1}$ \\
Field of View           & $1.3 \times 0.9$~deg$^{2}$ \\
\hline
Model                   & CMOS SONY IMX455 \\
Pixel Size              & $3.76\,\mu\mathrm{m}$ \\
Pixel Array Size (X, Y) & 9576 $\times$ 6388 \\
Readout Noise           & $3\,e^{-}$ \\
Dark Current            & $0.01\,e^{-}\,\mathrm{s}^{-1}$ \\
\hline
\multicolumn{2}{l}{\textit{Filters}} \\
Medium-band (20 filters) & m400, m425, m450, \dots, m875 \\
Medium-band (40 filters) & m400, m412, m425, \dots, m887 \\
Broad-band              & $u, g, r, i, z$ \\
\hline
\multicolumn{2}{l}{\textit{Observing Conditions}} \\
Exposures               & $100\,\mathrm{s} \times 3$ \\
Seeing                  & $2^{\prime\prime}$ \\
\enddata
\end{deluxetable}

\subsection{The 7-Dimensional Telescope and 7-Dimensional Sky Survey}\label{subsec:7dt}

\subsubsection{Hardware and System Overview} \label{sec:overview}

The 7-Dimensional Telescope (7DT) is a modular array of twenty 0.5-m telescopes in Chile. 
Each unit employs a high-resolution, fast-readout SONY IMX455 CMOS detector (9576\,$\times$\,6388; $\sim$60\,Mpix) with a pixel scale of $\sim$0.51\,arcsec\,pix$^{-1}$ and a 9-slot motorized filter wheel. 
As of August 2025, 16 of the planned 20 units are operational. 
We define two medium-band configurations used in this work: (1) the 40 filter set spanning $m400$–$m887$ in 12.5\,nm steps (the planned ultimate setup) and (2) the 20 filter set spanning $m400$–$m875$ in 25\,nm steps (current operational setup). 
Additional hardware specifications are provided in \citet{2024SPIE13094E..0XK}.


To acquire a single-epoch SED, the medium-band filters are partitioned across the individual telescopes, allowing the array to function as a single instrument that simultaneously captures photometric data in many different bands of the same target. 
By distributing the filters across multiple telescopes and stepping the wheels, the array completes the full medium-band set within minutes, effectively achieving an almost single-epoch spectro-photometric sampling of the optical sky. 
This observing configuration, referred to as the spectroscopic mode, produces a low-resolution ($R \sim 30$–$70$) SED over $\sim1.2\,\mathrm{deg}^2$, optimized for rapid transient classification (particularly KNe) without relying on multi-epoch color evolution.
For completeness, 7DT also supports a wide-search mode (uniform broad-band tiling for rapid areal coverage) and a deep mode (co-pointed exposures to increase depth); detailed use cases of these survey-oriented modes fall outside the scope of this work. 

7DT carries out the 7-Dimensional Sky  Survey (7DS), which is a wide-field, time-domain spectral mapping survey. 
7DS is composed of three surveys: Reference Image Survey (RIS), which spectrally maps the entire southern sky to use the image for Differential Image Analysis templates and complete low-resolution pictures, Wide-Field Time-Domain Survey (WTS) for weekly cadence monitoring of a wide area ($\sim 2,000$~\,deg$^2$) to trace the long-term variability of sources such as active galactic nuclei (AGNs), and Intensive Monitoring Survey (IMS) for daily monitoring of high-cadence monitoring studies (e.g., SNe, AGNs, TDEs, and solar moving objects) over a relatively small field of view ($\sim 10$~\,deg$^2$) for providing deep spectral map of the same area for extragalactic studies. 
Additionally, 7DS incorporates target-of-opportunity (ToO) observations of interesting transients such as GW events.



Building on this context, our study assumes the baseline 7DT observing sequence, and we construct simulated medium-band synthetic photometry to train the classification models (Table~\ref{tab:setting}). To emulate these observational conditions and evaluate the classifier performance in a controlled setting, we constructed a dedicated 7DT simulator described below.

\subsubsection{Synthetic Photometry} \label{sec:synphot}

Building on the instrumental characteristics described in Section~\ref{sec:overview}, we developed a dedicated simulator-\texttt{7DT-Simulator} \citep{paek2025_7dtsimulator}-to generate realistic synthetic photometry under the same conditions as the 7DT system, providing tools for estimating photometric depth and generating synthetic photometry from input spectra with tunable parameters.

To simulate realistic observational conditions, we convolved the filter response functions with the quantum efficiency (QE) of the CMOS sensor, atmospheric transmission from ESO \citep{2020SPIE11449E..1BB}, and the optical throughput of the system as listed in Table \ref{tab:setting} \citep{2024SPIE13094E..0XK,2025arXiv250922165K}.


To refine the optics efficiency model, we considered the specific configurations of the DR500 and CDK24 telescope systems. The total optical throughput was computed by combining internal transmittance values for optical materials (H-K9LA, H-LaK4L, and H-FK95N) and accounting for multiple passes through anti-reflection (AR)-coated elements and mirror reflectivities. 
The total efficiency is about 66.7\%, primarily.

The sky background model was generated with the \texttt{SkyCalc}\footnote{\url{https://www.eso.org/observing/etc/doc/skycalc/helpskycalc.html}} web application, which implements the Cerro Paranal Advanced Sky Model \citep{2012A&A...543A..92N,2013A&A...560A..91J}. We adopted typical conditions---airmass $\sim\!1.3$, grey time, and a Moon–target separation of $45^{\circ}$---and used the VLT Cerro Paranal site (altitude 2640 m) as a proxy because \texttt{SkyCalc} provides limited site customization. Under these assumptions, the sky brightness across the 7DT filter range is $\sim 20.5~\mathrm{mag~arcsec^{-2}}$. Details of the adopted sky model parameters are listed in Table~\,\ref{tab:setting}. 
However, since the 7DT site is located at an altitude of $\sim$1600~\,m our sky model may slightly overestimate sensitivity \citep{2024arXiv240616462K}.

\begin{figure*}
    \centering
    \includegraphics[width=0.75\linewidth]{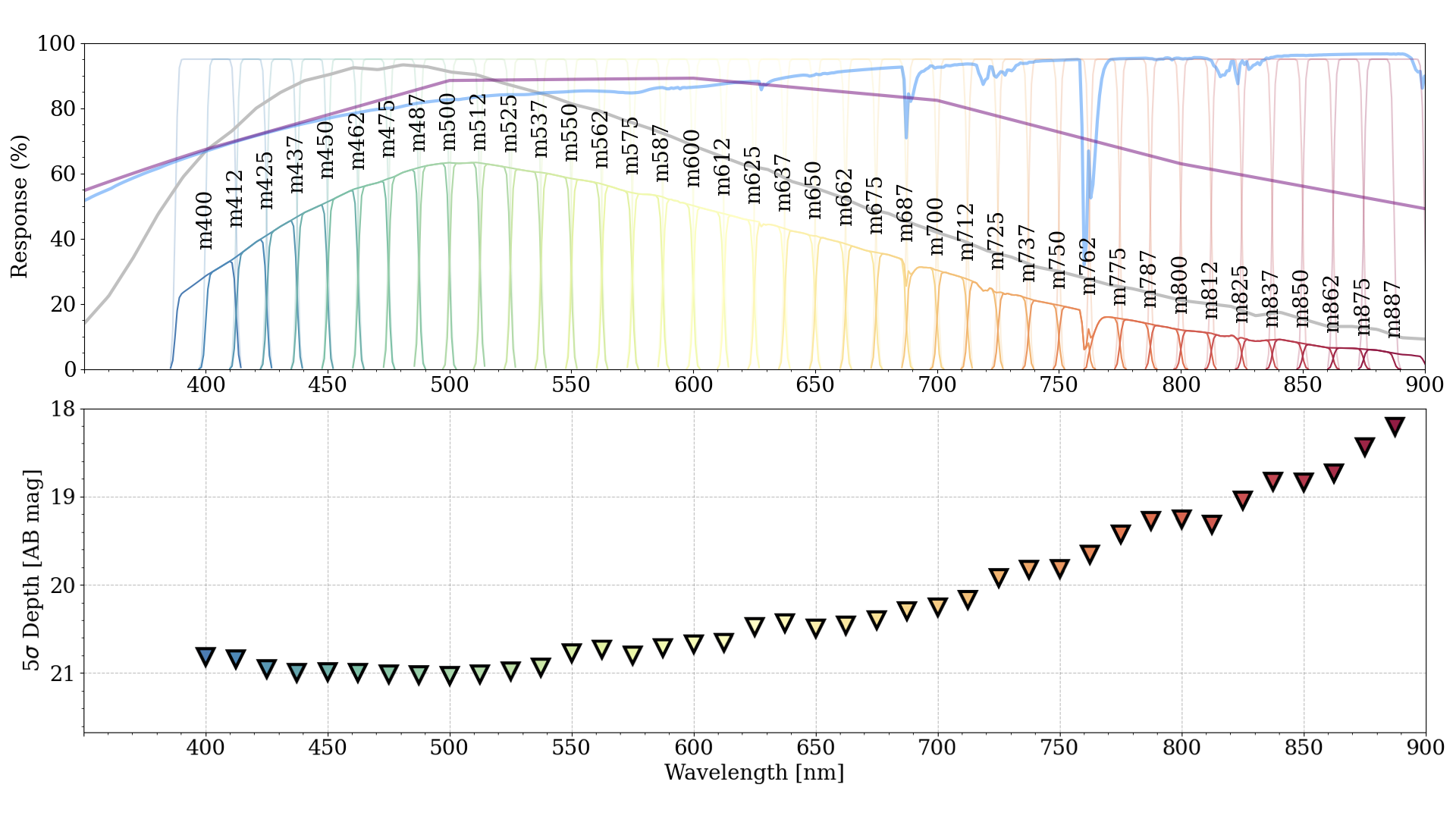}
    \caption{System response and $5\sigma$ point-source depth for the 7DT medium-band set, computed with \texttt{7DT-Simulator}. 
    \textbf{Top:} Light-colored curves show the nominal filter transmission profiles (top-hat-shaped). The gray, blue, and purple curves denote the detector QE (CMOS), atmospheric transmission, and optics throughput, respectively. Rainbow-colored curves are the final system responses (filter $\times$ optics $\times$ atmosphere $\times$ QE). 
    \textbf{Bottom:} Colored triangles mark the $5\sigma$ depths (AB mag) for individual medium bands.}
    \label{fig:response}
\end{figure*}

The transmission curves of 40 medium-band filters were modeled using top-hat functions. We adopted a default exposure time of $3 \times 100$ seconds and assumed a seeing value of 2\,$^{\prime\prime}$. The combined response curves--including the filter profiles, camera QE, atmospheric transmission, and optics efficiency--and their corresponding depths for each filter are shown in Figure~\ref{fig:response}.

Using the simulator, we generated synthetic photometric measurements with realistic observational scatter based on the calculated system sensitivity and instrumental settings. The synthetic photometry and the corresponding SNRs were computed through the following steps.

First, the photon rate was derived from the input spectrum's flux density $F_\nu(\lambda)$, wavelength $\lambda$, and the total system response $R(\lambda)$:

\begin{equation}
\text{(Photon Rate)} = \frac{\int \frac{R(\lambda) \cdot F_\nu(\lambda)}{\lambda} \, d\lambda}{h},
\end{equation}

where $h$ is Planck's constant.

Next, the effective observed flux was calculated as:

\begin{equation}
\text{(Observed Flux)} = \frac{\int \frac{R(\lambda) \cdot F_\nu(\lambda)}{\lambda} \, d\lambda}{\int \frac{R(\lambda)}{\lambda} \, d\lambda}.
\end{equation}

The SNR accounts for contributions from the source signal ($Q_{\rm src}$), sky background ($Q_{\rm sky}$), dark current ($Q_{\rm dark}$), and readout noise ($dQ_{\rm RN}$), as well as the number of pixels in the aperture ($N_{\rm aper}$):

\begin{equation}
\text{SNR} = \frac{Q_{\rm src}}{\sqrt{Q_{\rm src} + N_{\rm aper} (Q_{\rm sky} + Q_{\rm dark} + dQ_{\rm RN}^2)}}.
\end{equation}

The corresponding magnitude error ($\sigma_m$) is expressed as:



\begin{equation}
\sigma_m = 2.5 \log_{10}\!\left(1 + \frac{1}{{\rm SNR}}\right)
\label{eq:magerr_from_snr}
\end{equation}

An observed magnitude is then drawn as a Gaussian realization around the true value:

\begin{equation}
m_{\rm obs} \sim \mathcal{N}\!\left(m,\; \sigma_m^2\right),
\label{eq:mag_realization}
\end{equation}

where the sampling is effectively limited to $[\,m-10\sigma_m,\; m+10\sigma_m\,]$.

Finally, we include a filter-dependent zeropoint error $\sigma_{\rm zp}$, adopting a representative value of 0.01~mag (corresponding to $\sim$1\%), and report the total measured uncertainty as the quadrature sum:

\begin{equation}
\sigma_{m,{\rm obs}} = \sqrt{\sigma_m^2 + \sigma_{\rm zp}^2}.
\label{eq:total_magerr}
\end{equation}

\begin{figure*}
    \centering
    \includegraphics[width=0.75\textwidth]{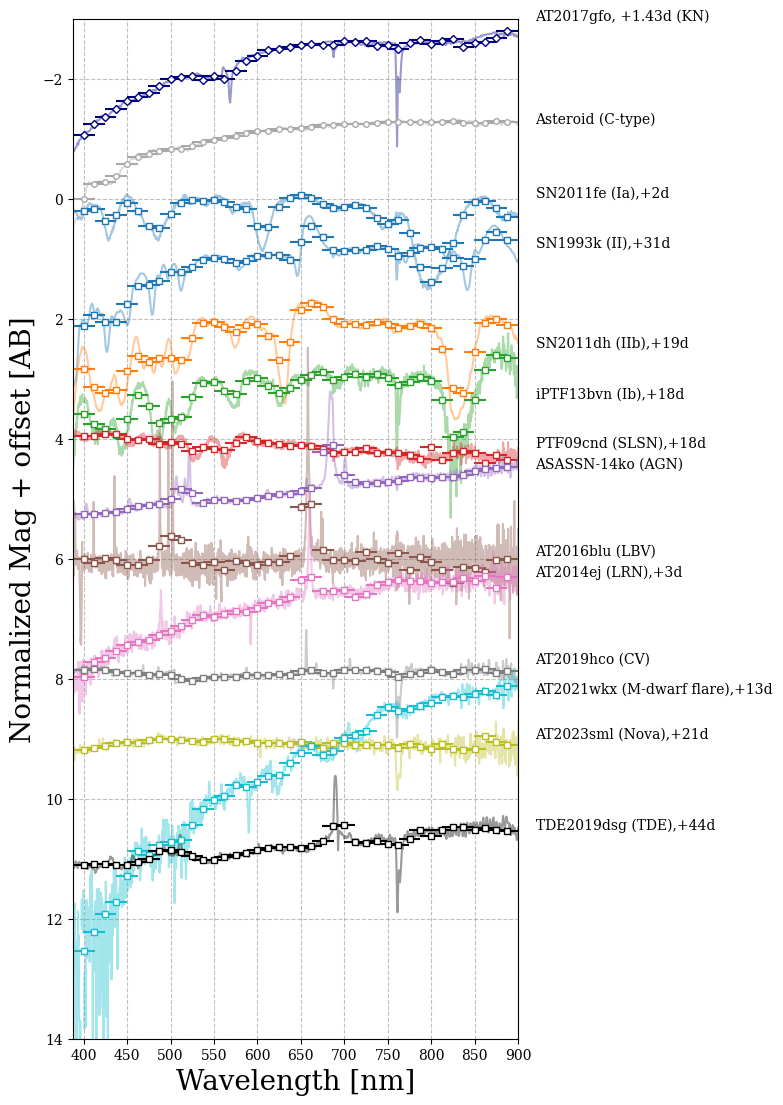}
    \caption{
    Examples of synthetic photometry (markers with connecting lines) and corresponding spectra generated by the \texttt{7DT-Simulator} using the 40 medium-band filter set of 7DT. 
    Each spectrum is normalized to zero magnitude at 550\,nm and vertically offset and denoted their name, type, approximate phase after discovery date.
    The KN spectrum is computed from the ENGRAVE observation of AT~2017gfo \citep{2022MNRAS.515..631G,2024MNRAS.529.2918G}, while the asteroid spectrum corresponds to a C-type object based on the Bus--DeMeo taxonomy template \citep{2009Icar..202..160D}. 
    All other transients are derived from spectral templates obtained via the OSC or WISeREP. 
    For each transient type, one representative spectrum was randomly selected from the available templates to visually illustrate its characteristic spectral features.
    }
    \label{fig:synphot_example}
\end{figure*}

Figure~\ref{fig:synphot_example} illustrates examples of synthetic photometry for diverse transient classes generated by the \texttt{7DT-Simulator} using the 40 medium-band filters. The medium-band SEDs not only reproduce the global spectral shapes of each class but also capture strong and wide emission and absorption features with sufficient fidelity to reveal their diagnostic differences. The corresponding spectra compiled for this figure are described in next Section~\ref{sec:spectra}.

This study is designed to reflect realistic 7DT operating conditions. Except for the asteroid class, all training data originate from observational spectra, capturing the diversity of real transients. Mock 7DT observations were generated with the \texttt{7DT-Simulator}, which includes telescope optics, wavelength-dependent filter sensitivity, and other observing parameters. Detection thresholds were enforced by requiring $5\sigma$ significance in at least one filter, and non-detections were explicitly encoded as \texttt{NaN} values. We note, however, that the current sky-background model is adopted from the VLT, which is at higher altitude than 7DT; this may mildly overestimate sensitivity relative to on-site conditions.

\begin{deluxetable*}{cccccc}
    \tablecaption{Mapping from Initial Claimed Type to Broad Type, with Data Source, Selected Counts, and Augmentation/Detection Statistics\label{tab:type_mapping}}
    \tablewidth{0pt}
    \tablehead{
      \colhead{Initial Claimed Type} & \colhead{Label} & \colhead{Data Source} & \colhead{Selected Count} & \colhead{Augmented Number} & \colhead{Detected Number}
    }
    \startdata
    Computed-Ia, Ia CSM, Ia Pe, Ia Pec  & Ia                 & OSC      & 2838 & 11352 & 10976 \\
    Ia-, Ia-02c, Ia-02cx, Ia-SC         &                    &          &      &       &       \\
    Ia-91T, Ia-91b, Ia-91bg, Iax[02     &                    &          &      &       &       \\
    \midrule
    II, II , II/, II?, II Pec, IIP      & II                 & OSC      & 1856 & 11136 & 11022 \\
    IIb, IIn/, IIn/LBV, IIn Pec         &                    &          &      &       &       \\
    \midrule
    Ib, Ib Pec, Ib-Ca, Ib/c, Ibn        & Ibc                & OSC      & 357  & 10353 & 10150 \\
    Ic, Ic BL, Ic Pec, Ic-hy, IIb/Ib/Ic &                    &          &      &       &       \\
    \midrule
    SL, SLS, SLSN                       & SLSN               & WISeREP  & 82   & 10004 & 9150  \\
    SLSN-I, SLSN-II, SLSN-R             &                    &          &      &       &       \\
    \midrule
    AGN                                 & AGN                & WISeREP  & 160  & 10080 & 9891  \\
    \midrule
    TDE                                 & TDE                & WISeREP  & 372  & 10044 & 10044 \\
    \midrule
    CV                                  & CV (SV)            & WISeREP  & 331  & 10032 & 9790  \\
    LBV                                 & LBV (SV)           &          & 38   &       &       \\
    LRN                                 & LRN (SV)           &          & 29   &       &       \\
    Nova                                & Nova (SV)          &          & 40   &       &       \\
    M dwarf                             & M-dwarf flare (SV) &          & 15   &       &       \\
    \midrule
                                       & Asteroid           & \citealt{2009Icar..202..160D} &     & 10000 & 9582  \\
    \midrule
                                       & Total              &          & 7654 & 82997 & 81705 \\
    \enddata
    \tablecomments{
        The table summarizes how initial claimed types in the source catalogs were consolidated into the broad classification categories adopted in this study. 
        The \emph{Data Source} column indicates whether spectra were obtained from the Open Supernova Catalog (OSC; \citealt{2017ApJ...835...64G})\protect\footnote{\url{https://github.com/astrocatalogs/SNe}}, from WISeREP (\citealt{2012PASP..124..668Y}), or from the Bus--DeMeo asteroid taxonomy \citep{2009Icar..202..160D}. 
        The stellar-variable (SV) label includes five subclasses: CV, LBV, LRN, nova, and M-dwarf flares.
        The \emph{Selected Count} gives the number of spectra that pass our quality cuts, requiring wavelength coverage of 387.5–900\,nm and median flux $<10^{-10}$. 
        The \emph{Augmented Number} represents the expanded sample size after applying data augmentation (see Section~\ref{sec:method}), and the \emph{Detected Number} shows how many instances of each class were ultimately identified by the classifier on the test set.
    }
\end{deluxetable*}

\subsection{Training and Validation Data} \label{sec:spectra}


\subsubsection{Archival non-KN Spectra} \label{subsubsec:nonkn}

To build a training sample for both the multi-class and anomaly classifiers, we collected archival spectra of diverse transient types from public databases.
The selection of transient types was guided by three factors: (1) their occurrence in wide-field transient searches and GW follow-up campaigns, (2) their potential to act as KN impostors in photometric classification, and (3) the scientific goals of the 7DT survey.

Our dataset includes eight broad transient categories: Type Ia, Ibc, and II SNe, superluminous SNe (SLSNe), tidal disruption events (TDEs), active galactic nuclei (AGN), asteroids, and a diverse group of stellar variables (SVs), including cataclysmic variables (CVs), luminous blue variables (LBVs), luminous red novae (LRNs), novae, and M-dwarf flares. Type Ia, Ibc, and II SNe were compiled from the \textbf{Open Supernova Catalog} (OSC; \citealt{2017ApJ...835...64G})\footnote{\url{https://github.com/astrocatalogs/SNe}}. From the total of 20,821 spectra in OSC, we selected 5,311 (25.5\%) using two criteria: (1) wavelength coverage spanning at least 387.5–900\,nm, and (2) median flux $<10^{-10}$ (cgs) for entries whose metadata listed flux units as \texttt{erg cm$^{-2}$ s$^{-1}$ \AA$^{-1}$}, given occasional unit inconsistencies. All other transient types (SLSNe, TDEs, AGN, CVs, LBVs, novae, and M-dwarf flares) were obtained from the \textbf{Weizmann Interactive Supernova REpository} (WISeREP; \citealt{2012PASP..124..668Y}), queried on 2025-05-10 (HST). We deliberately avoided overly fine-grained subtyping. Instead, we mapped all available subtypes into broader classes (Table~\ref{tab:type_mapping}) to better reflect the achievable discrimination power at $R\sim$30–70.

Briefly, Ia denote thermonuclear white-dwarf explosions; Ibc/II are core-collapse SNe (stripped-envelope and H-rich, respectively); SLSNe are superluminous core-collapse events; TDEs arise from stellar disruption by supermassive black holes; AGN are accreting supermassive black holes with stochastic variability; SVs encompass eruptive/accretion-driven stellar variability (e.g., CVs, LBVs, novae, M-dwarf flares); and asteroids exhibit reflected-solar continua with weak compositional slopes. The SV category groups multiple stellar-variability phenomena rather than treating each as a separate class. Although SVs are not primary 7DT/7DS science targets, they are important as potential KN contaminants to efficiently identify and rule out. LBVs are of particular interest, having been repeatedly reported as one of the major contaminants in optical KN searches \citep{2025MNRAS.542..541F}; CVs, novae, and M-dwarf flares \citep{2020MNRAS.491...39C} can also mimic fast, KN-like photometric behavior.

The spectral data are heterogeneous in quality, format, and metadata consistency because OSC and WISeREP aggregate observations from a wide variety of instruments and pipelines, so careful selection of clean and appropriate data is required. Accordingly, from 1,240 spectra spanning 10 broad types, we imposed a uniform wavelength-coverage criterion (387.5–900\,nm) and retained only those that additionally passed the median-flux cut when applicable. Because flux-calibration metadata in the archives are often unreliable---for example, recorded flux units sometimes do not match the actual spectrum scaling---we required physically plausible scaling in units of erg\,s$^{-1}$\,cm$^{-2}$\,\AA$^{-1}$, keeping only spectra with $\log_{10}(\mathrm{median\ flux})<-10$; We then conducted a by-eye quality screen of every spectrum and removed those that were too noisy or unnaturally featureless; objects without a secure classification were also discarded.

\begin{figure*}
    \centering
    \includegraphics[width=1\linewidth]{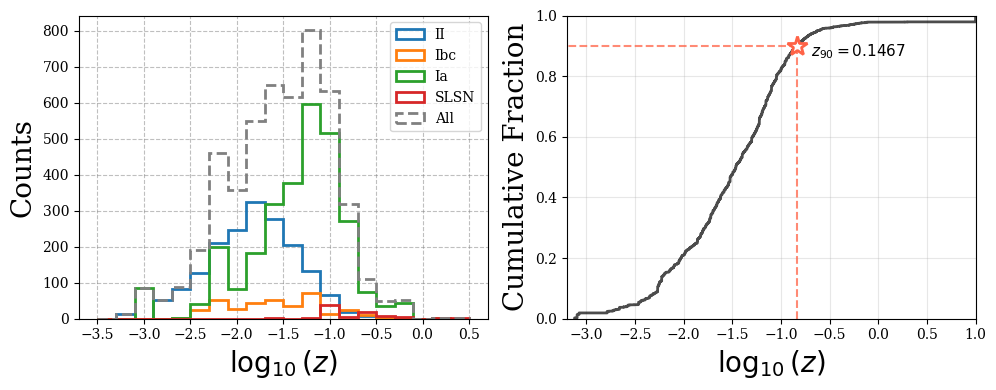}
    \caption{
Redshift distributions for SN subclasses (Type~II, Type~Ibc, Type~Ia, and SLSN) with valid redshift metadata. 
\textbf{Left:} Class-wise histograms of $\log_{10}(z)$; the dashed gray line shows the distribution for the combined sample. 
Only objects with usable redshift entries were included in each histogram. 
\textbf{Right:} Empirical cumulative distribution of $\log_{10}(z)$ for the same sample. 
The red star marks the 90th percentile ($z_{90}\approx0.1467$), indicating that $\sim$90\% of the usable sample lies at $z<0.15$.
}
    \label{fig:redshift}
\end{figure*}

We constructed redshift distributions using only entries with existing redshift metadata from the compiled archival sources; these are shown in Figure~\ref{fig:redshift}. 
We include only spectra with a recorded redshift; consequently, the histograms comprise the SN subclasses (Types~Ia, Ibc, II, and SLSNe). 
Non-SN classes (e.g., AGN, TDE, CV, LBV, novae, and M dwarf flares) generally lack reliable redshift measurements in our compilation and are therefore not shown.
Consequently, the resulting distributions are observationally biased toward nearby events: the cumulative distribution indicates that about $90\%$ of the usable non--KN sample lies at $z<0.15$.


Throughout this paper, we adopt shortened notations for three SN types (Ia, Ibc, II) as defined in Table~\ref{tab:type_mapping}; these abbreviations are used consistently unless otherwise noted.

Asteroids are frequently detected as foreground contaminants during wide-field extragalactic transient surveys and are also one of the main science interests of 7DS. We generated a set of synthetic asteroid spectra based on the Bus--DeMeo taxonomy \citep{2009Icar..202..160D}. A complete description of the construction procedure, including reflectance extrapolation, solar spectrum multiplication, flux scaling, and metadata compilation, is provided
in Appendix~\ref{app:asteroid}.

\subsubsection{Simulated KN Spectra}\label{subsubsec:simkn}

In this study, we define ``anomaly'' not as an entirely unknown or theoretically speculative phenomenon, but as a rare transient class for which limited observational data exist, preventing robust supervised training. KNe fall into this category because, while they have been observed once spectroscopically (AT~2017gfo), they remain exceptionally rare in current surveys. 

To generate KN spectra considering various physical parameters as a test sample, including ejecta properties and geometry, we choose the two-component, axisymmetric radiative-transfer grid of \citet{2021ApJ...918...10W}. 
This model decomposes the ejecta into a low-$Y_e$ (lanthanide-rich) dynamical component and a higher-$Y_e$ (lanthanide-poor) wind component with distinct morphologies. 
We use these model spectra to evaluate our KN anomaly-detection model as a function of key physical parameters and to calibrate its operating thresholds accordingly.
In this work, we fix the morphology to the AT~2017gfo-like configuration provided in \citet{2021ApJ...918...10W}-namely, a toroidal dynamical ejecta plus a ``peanut''-shaped high-$Y_e$ wind (T+P)-and vary the bulk masses and velocities of the two components. 






Starting from the discrete grid of KN model, we construct a regular, high-dimensional cube via multi-linear interpolation so that spectra can be sampled not only at native grid nodes but also at intermediate parameter values. 
We denote \(m_{\mathrm d}\) and \(v_{\mathrm d}\) as the dynamical–ejecta mass and velocity, \(m_{\mathrm w}\) and \(v_{\mathrm w}\) as the wind–ejecta mass and velocity, \(\theta\) as the viewing angle, and \(t\) as the phase (days post-merger). 
Specifically, we adopt
\begin{equation}
m_{\mathrm d} \in \{0.001,\,0.003,\,0.01,\,0.03,\,0.1\}\,M_\odot
\end{equation}
\begin{equation}
v_{\mathrm d} \in \{0.05,\,0.15,\,0.3\}\,c
\end{equation}
\begin{equation}
m_{\mathrm w} \in \{0.001,\,0.003,\,0.01,\,0.03,\,0.1\}\,M_\odot
\end{equation}
\begin{equation}
v_{\mathrm w} \in \{0.05,\,0.15,\,0.3\}\,c
\end{equation}
\begin{equation}
\theta \in \{0^\circ,\,15^\circ,\,30^\circ,\,45^\circ,\,60^\circ,\,75^\circ,\,90^\circ\}
\end{equation}
\begin{equation}
t \in \{0.125,\,0.25,\,0.5,\,1.0,\,2.0,\,3.0\}\ \text{days post-merger}
\end{equation}

Spectra are evaluated on a wavelength grid from 3000 to 12{,}000~\AA\ with 5~\AA\ spacing and scaled to a fiducial luminosity distance of $D_L \simeq 40$~Mpc ($z \simeq 0.010$).  
This parameterization yields $5\times3\times5\times3\times7\times6=9450$ spectra.

\begin{figure*}
    \centering
    \includegraphics[width=1\linewidth]{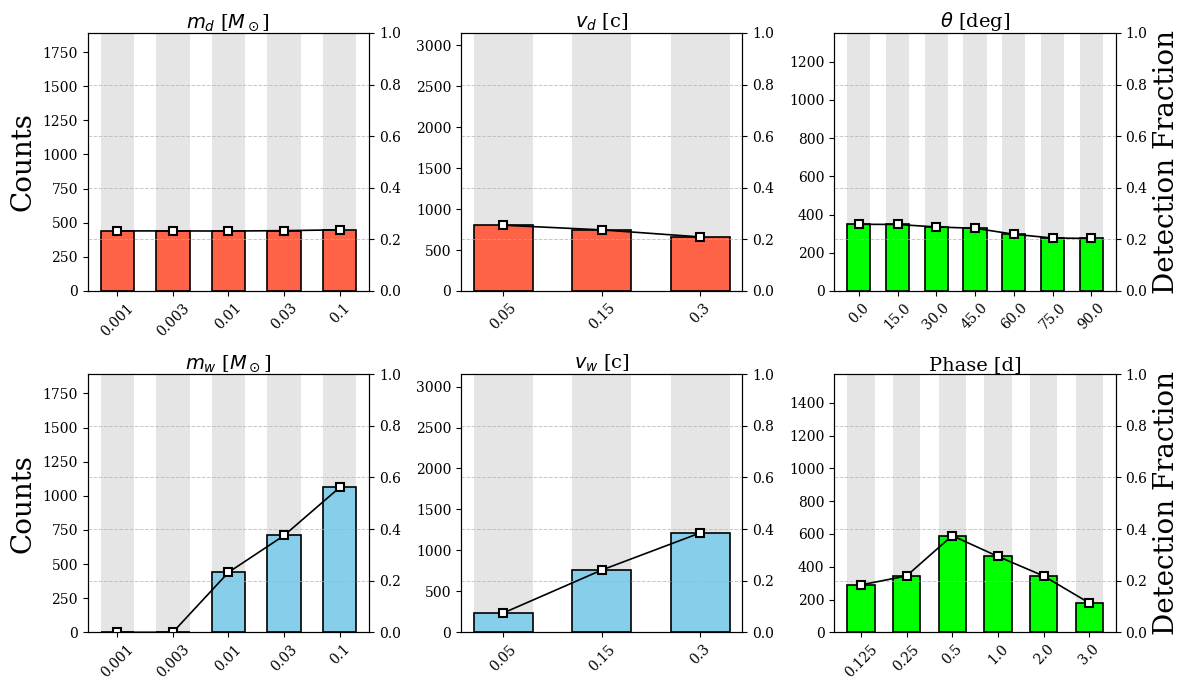}
    \caption{
    Detection efficiency of simulated KNe with the 7DT as a function of model parameters. 
    Bars show the number of simulated samples per parameter bin (left axis), while black squares and lines indicate the corresponding detection fractions (right axis). 
    A KN is considered detected if at least one medium-band filter achieves a $5\sigma$ detection in the simulated 7DT observation. 
    The parameters shown are the dynamical ejecta mass ($m_{d}$), dynamical ejecta velocity ($v_{d}$), wind ejecta mass ($m_{w}$), wind ejecta velocity ($v_{w}$), viewing angle ($\theta$), and phase after merger. 
    Detection fractions are computed only for simulated KN that are detectable under the assumed 7DT sensitivity.
    %
    }
    \label{fig:kn_detect_frac}
\end{figure*}


We further examined which types of simulated KNe are preferentially detected by the 7DT (Figure~\ref{fig:kn_detect_frac}). 
Here, we define a detection as having $\mathrm{SNR}>5$ in at least one medium-band filter. Out of all simulated KN spectra, 2209 (23.4\%) have peak brightness within 7DT detection limits, indicating that the majority of KN models would be too faint to detect in practice. 
The detection efficiency is strongly dependent on the wind ejecta properties: larger $m_{w}$ and higher $v_{w}$ values produce brighter optical emission and result in higher detection fractions. 
By contrast, the $m_{d}$, $v_{d}$, and $\theta$ show relatively uniform detection probabilities, indicating that their influence on the observable optical brightness is less critical for 7DT sensitivity. 
The phase dependence reveals a clear peak at $\sim$0.5\,d post-merger, consistent with the expected maximum in the blue optical bands, after which the detectability declines rapidly. 
This suggests that the 7DT is naturally biased toward identifying KNe with luminous, rapidly evolving optical components, and that early-time follow-up is particularly crucial for maximizing detection yield.
Because only the simulated KNe that are successfully detected enter our test sample, this selection inherently biases the evaluation toward intrinsically bright cases, particularly those with high wind ejecta masses and velocities. Consequently, the classifier's performance on faint KNe remains unconstrained and may be systematically optimistic when applied to the full KN population.

\subsubsection{AT~2017gfo KN Spectra} \label{subsubsec:engrave}

We adopt the publicly released spectroscopic sequence of AT~2017gfo from the ENGRAVE collaboration\footnote{\url{https://www.engrave-eso.org/AT~2017gfo-Data-Release/}} as the test sample for anomaly detection model, obtained with the ESO VLT/X-shooter instrument. The dataset provides nightly coverage for seven consecutive days (from +1.43\,d to +7.4\,d after merger) with complete wavelength coverage from $\sim$3300\,\AA\ to 2.5\,$\mu$m, combining the UVB, VIS, and NIR arms. All spectra were uniformly reduced with the ENGRAVE pipeline \citep{2022MNRAS.515..631G,2024MNRAS.529.2918G} and spectro-photometrically calibrated to broad-band photometry, and are recommended as a definitive reference set for modeling AT~2017gfo.

For consistency with 7DT observing conditions, we used the telluric– and redshift–uncorrected versions so as to retain instrumental/atmospheric features that would realistically appear in ground-based follow-up. 

In the UVB–VIS overlap (5300–5600\,\AA), we interpolated masked regions using a local-trend method. 
We first estimated a robust continuum by coarse binning (100\,\AA), then fit simple linear trends to the adjacent segments on either side of each masked gap (within $\pm$300–500\,\AA). 
The two trends were blended across the gap, resampled onto a uniform 10\,\AA\ grid, and merged with the unmasked data.



\begin{figure*}
    \centering
    \includegraphics[width=0.75\linewidth]{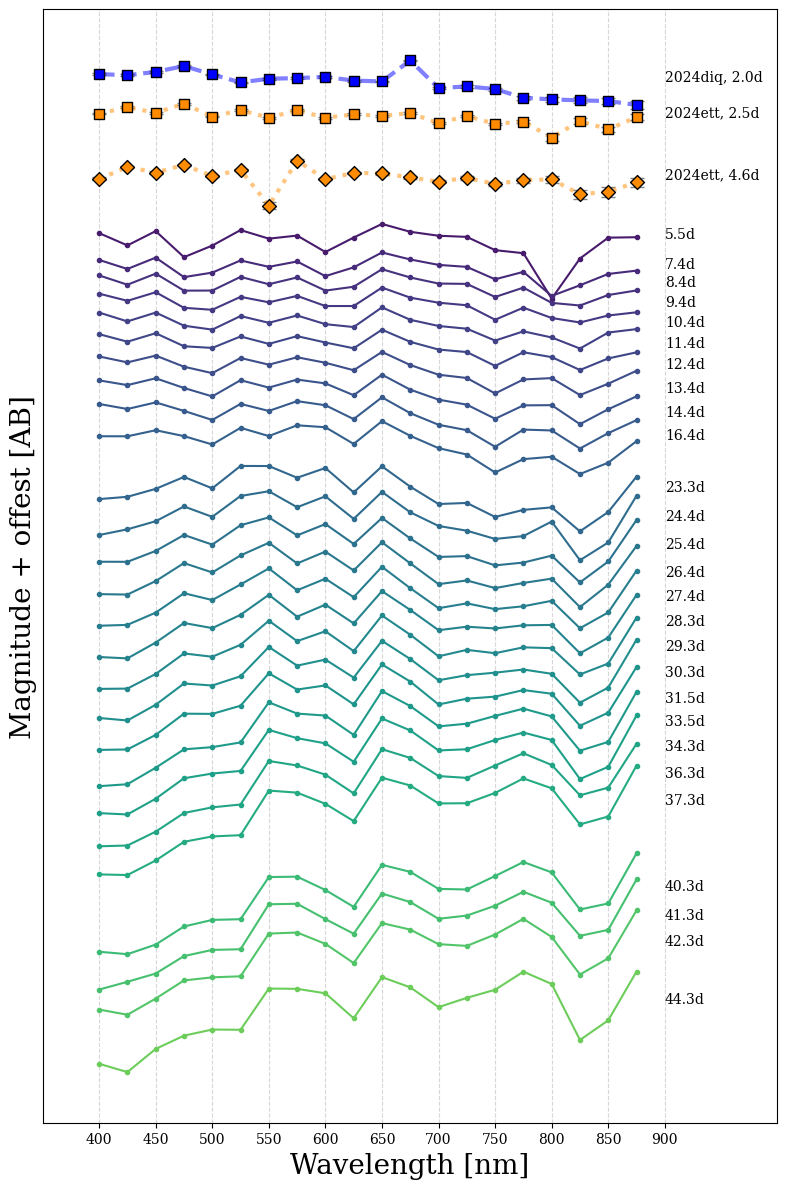}
    \caption{
Transient SEDs observed by 7DT in 20 filter set.
The stacked curves at the bottom trace the temporal evolution of the Type~Ia~SN, SN~2025fvw over
$\sim$3–42\,days post-discovery (2025 March 26 19{:}09{:}59; \citealt{2025TNSTR1151....1I});
points are shown with dot markers connected by solid lines, and magnitudes are vertically
offset for clarity (AB system; the $x$-axis is the filter central wavelength in nm).
The three curves at the top are single-epoch SEDs for real test samples:
the Type~II~SN, SN~2024diq at 2.0\,days post-discovery (2024 February 28 08{:}09{:}46.368;
\citealt{2024TNSCR.615....1C,2024TNSCR.614....1D}), plotted with square markers connected by
dashed lines; and the CV, AT~2024ett at 2.5 and 4.6\,days post-discovery
(2024 March 19 11{:}40{:}04.224; \citealt{2024TNSCR.896....1D}), plotted with square (2.5\,d) and
diamond (4.6\,d) markers connected by dotted lines.
The corresponding photometry is summarized in Table~\ref{tab:7dt_val}.
}
    \label{fig:7dtobs_all}
\end{figure*}

\subsection{Observational Test Sample}\label{subsec:7dt_test}
To verify real-world performance, we used a test set observed with 7DT (20-filter configuration; 25\,nm spacing). Each visit consisted of $3\times100$\,s exposures. Test sets include multi-epoch SEDs of SN~2025fvw (Type~Ia SN; \citealt{2025TNSCR1320....1F,2025TNSCR1262....1L,2025TNSCR1186....1R}), a single epoch SED of SN~2024diq (Type~II SN; \citealt{2024TNSCR.615....1C,2024TNSCR.614....1D}), and multi-epoch SEDs of AT~2024ett (CV; \citealt{2024TNSCR.896....1D}) as described in Table \ref{tab:7dt_val}. Standard calibrations (bias, dark, flat), astrometric/photometric calibration, and aperture photometry were applied with the gpu-accelerated pipeline (\texttt{gpPy-GPU}; \citealt{paek2025_gppy_gpu}). Table~\ref{tab:7dt_val} summarizes the observed targets and Figure~\ref{fig:7dtobs_all} shows SEDs of each transient; these data are used for test of trained models in Section~\ref{sec:results}.

\begin{deluxetable*}{lccccc}
    \tablecaption{Classification results for the 7DT test sample: multi-class predictions with scores and anomaly scores\label{tab:7dt_val}}
    \tablehead{
        \colhead{Object (Type)} & \colhead{This Study} &
        \colhead{$\rm 1-P_{\max,20}$} & \colhead{$\rm P_{\mathrm{ano}}$} &
        \colhead{MJD} & \colhead{$\Delta t$ (days)}
    }
    \startdata
        SN~2025fvw (Ia)\tablenotemark{1} & Ia & 0.005 & -0.090 & 60763.1 & 3.08 \\
                                         & Ia & 0.003 & -0.133 & 60765.0 & 5.04 \\
                                         & Ia & 0.003 & -0.133 & 60766.0 & 6.04 \\
                                         & Ia & 0.005 & -0.136 & 60767.0 & 6.98 \\
                                         & Ia & 0.002 & -0.131 & 60768.0 & 7.98 \\
                                         & Ia & 0.002 & -0.122 & 60769.0 & 8.98 \\
                                         & Ia & 0.001 & -0.110 & 60770.0 & 9.97 \\
                                         & Ia & 0.002 & -0.107 & 60771.0 & 10.97 \\
                                         & Ia & 0.002 & -0.109 & 60772.0 & 11.96 \\
                                         & Ia & 0.003 & -0.081 & 60774.0 & 13.96 \\
                                         & Ia & 0.004 & -0.075 & 60780.9 & 20.94 \\
                                         & Ia & 0.005 & -0.101 & 60782.0 & 22.00 \\
                                         & Ia & 0.005 & -0.105 & 60783.0 & 22.96 \\
                                         & Ia & 0.006 & -0.111 & 60784.0 & 23.96 \\
                                         & Ia & 0.007 & -0.116 & 60785.0 & 24.95 \\
                                         & Ia & 0.011 & -0.112 & 60786.0 & 25.95 \\
                                         & Ia & 0.010 & -0.113 & 60786.9 & 26.94 \\
                                         & Ia & 0.018 & -0.119 & 60787.9 & 27.94 \\
                                         & Ia & 0.015 & -0.100 & 60789.1 & 29.09 \\
                                         & Ia & 0.009 & -0.095 & 60791.1 & 31.07 \\
                                         & Ia & 0.007 & -0.086 & 60791.9 & 31.93 \\
                                         & Ia & 0.011 & -0.082 & 60793.9 & 33.92 \\
                                         & Ia & 0.023 & -0.086 & 60794.9 & 34.92 \\
                                         & Ia & 0.019 & -0.081 & 60797.9 & 37.91 \\
                                         & Ia & 0.024 & -0.085 & 60798.9 & 38.91 \\
                                         & Ia & 0.015 & -0.086 & 60799.9 & 39.91 \\
                                         & Ia & 0.015 & -0.086 & 60801.9 & 41.91 \\
        \midrule
        SN~2024diq (II)\tablenotemark{2} & TDE & 0.232 & -0.117 & 60781.3 & 1.97 \\
        \midrule
        AT~2024ett (CV; SV)\tablenotemark{3} & Ia  & 0.254 & -0.139 & 60388.5 & 2.54 \\
        AT~2024ett (CV; SV)\tablenotemark{3} & Ia  & 0.306 & -0.110 & 60390.6 & 4.62 \\
    \enddata
    \tablenotetext{1}{Type reference: \citep{2025TNSCR1320....1F,2025TNSCR1262....1L,2025TNSCR1186....1R}. $\Delta t$ values were calculated from the discovery date, 2025 March 26 19:09:59.}
    \tablenotetext{2}{Type reference: \citep{2024TNSCR.615....1C,2024TNSCR.614....1D}. $\Delta t$ was calculated using the discovery date, 2024 February 28 08:09:46.368.}
    \tablenotetext{3}{Type reference: \citep{2024TNSCR.896....1D}. $\Delta t$ values were calculated from 2024 March 19 11:40:04.224.}
    \tablecomments{All observations used the 20-filter set. Photometry and quality cuts match Section~\ref{sec:synphot}.}
\end{deluxetable*}

\section{Methods} \label{sec:method}

\subsection{Feature Construction and Preprocessing}




To enhance the discriminative power of the classifier and not be biased by the observed flux and focus on the spectral shape and features while maintaining interpretability, we used colors as input features based on the SEDs sampled by the 7DT medium-band filters. These features were constructed from the observed magnitudes in the 20 and 40 filter sets for each. 
All possible pairwise color indices, defined as the magnitude differences between two filters (for example, $\rm m_{m400} - m_{m412}$ and $\rm m_{m600} - m_{m650}$), were computed for each filter set. 
These features capture not only the spectral slopes but also broad emission or absorption lines and improve class separability by encoding relative brightness. As a result, we got 190 (780) color features for 20 (40) filter sets corresponding to $N_{\rm filter}(N_{\rm filter}-1)/2$ combinations of filters, where $N_{\rm filter}$ is the number of filters.

This approach allows the ML models to learn under realistic detection contexts, where some filters may not yield a measurement. We compute features only for sources with at least one band satisfying \(\mathrm{SNR}>5\) (our detection criterion), thereby restricting the analysis to realistically detectable cases. Photometric points with \(\mathrm{SNR}<3\) are treated as non-detections, assigned \texttt{NaN} values, and any color features involving those bands are masked. 

%


\subsection{Data Balancing and Augmentation}

As mentioned in Section \ref{sec:data}, the transient classes are highly imbalanced in their original distributions. To mitigate this and to increase the diversity and realism of the training set, we performed data augmentation using the \texttt{7DT-Simulator}. The simulator adds random Gaussian-distributed noise consistent with the system sensitivity, thereby ensuring that the augmented samples better reflect realistic observing conditions. To balance the classes, we increased the number of synthetic photometry samples so that each class reached approximately $10^4$ instances. Specifically, we scaled the original selected counts by integer multipliers: AGN (160 $\times$ 63 = 10,080), SN II (1856 $\times$ 6 = 11,136), SN Ia (2838 $\times$ 4 = 11,352), SN Ibc (357 $\times$ 29 = 10,353), SLSNe (82 $\times$ 122 = 10,004), stellar variables (SV; 456 $\times$ 22 = 10,032, including CVs, LBVs, LRNs, novae, and M-dwarf flares), and TDEs (372 $\times$ 27 = 10,044) as listed in Table~\ref{tab:type_mapping}.

\subsection{The Hybrid Classification Framework}

We train two model families: (i) a supervised multi-class classifier for eight non-KN categories (Ia, Ibc, II, SLSN, AGN, TDE, SV, asteroids), and (ii) an unsupervised anomaly detector targeting KNe.

We trained models using two filter set configurations as mentioned in Section~\ref{subsec:7dt}: 
Synthetic photometry was generated as described in Section~\ref{sec:synphot}, and class imbalance was addressed via sample augmentation (see Table~\ref{tab:type_mapping}). A train–test split of 80\%/20\% was applied, with care taken to ensure that augmented duplicates remained within the same partition.

\subsubsection{Multi-class Classifier}\label{subsubsec:multiclassclassifier}

\begin{figure}
    \centering
    \includegraphics[width=1\linewidth]{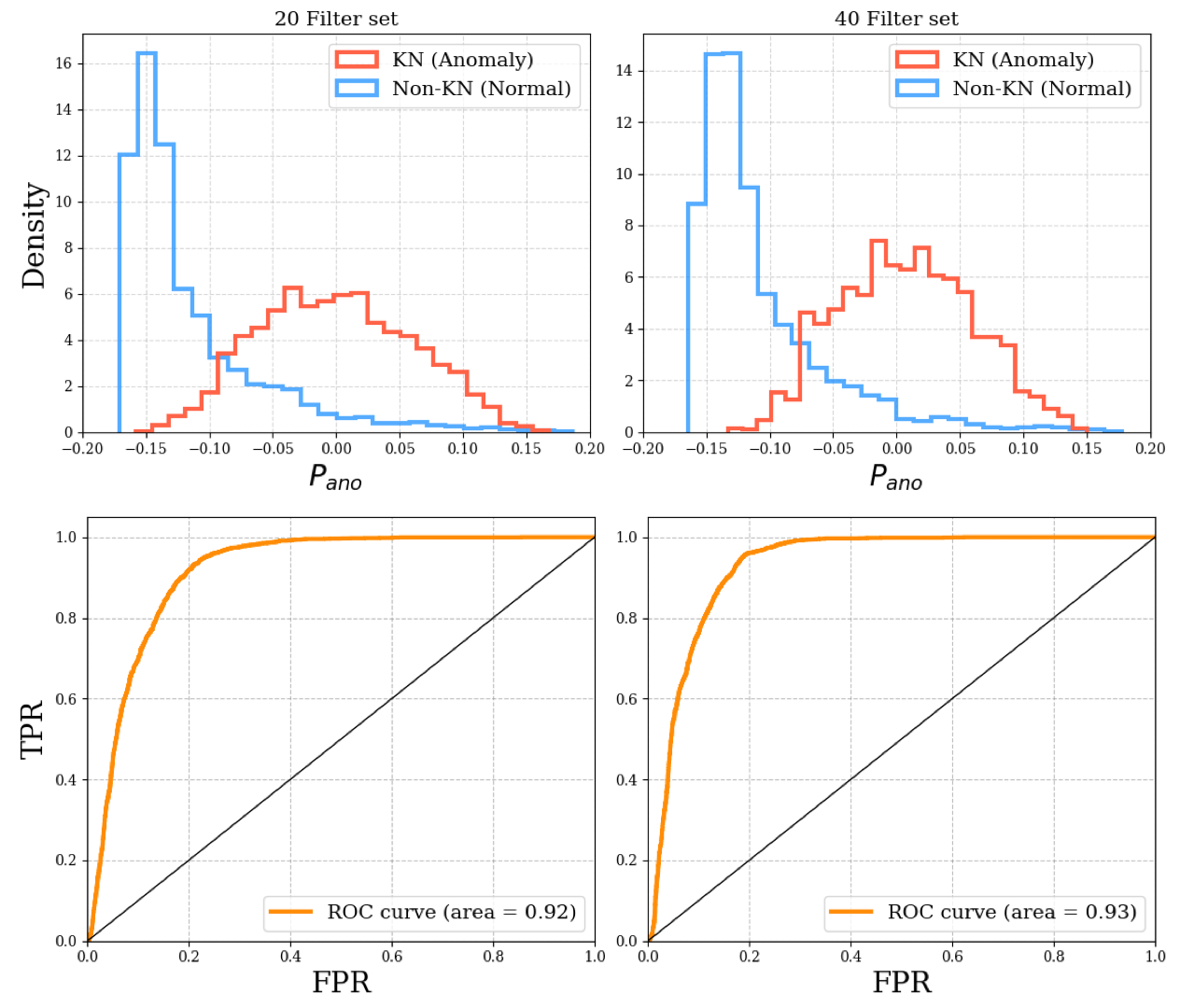}
    \caption{
    \texttt{iForest}'s anomaly classification results for KN identification. 
    Top row panels (left: 20 filter set, right: 40 filter set) show the distributions of the anomaly score $P_{\mathrm{ano}}$, with red histograms for KNe (anomalies) and blue histograms for non-KN transients (normal classes). 
    Bottom panels present the corresponding receiver operating characteristic (ROC) curves, yielding area under the curve (AUC) values of 0.92 for the 20-filter configuration (left) and 0.93 for the 40-filter configuration (right). 
    }
    \label{fig:iforest}
\end{figure}

To select a \emph{base model} for the multi-class classifier, we benchmark three gradient-boosted decision-tree (GBDT) implementations---\texttt{LightGBM} (Light Gradient Boosting Machine; \citealt{10.5555/3294996.3295074}), \texttt{CatBoost} \citep{2018arXiv181011363V}, and \texttt{XGBoost} (eXtreme Gradient Boosting; \citealt{2016arXiv160302754C}). GBDTs are tree-ensemble learners trained via gradient boosting and are well known to excel on tabular features (e.g., our medium-band colors), with strong performance in diverse benchmarks such as Kaggle\footnote{\url{https://www.kaggle.com}} \citep{2018ITVCG..24..163L}. While all three share core strengths in handling tabular data, they differ in optimization, feature handling, and missing value processing---each employing intrinsic mechanisms to handle masked non-detections (e.g., $\mathrm{SNR}<3$), thereby treating absent fluxes not as lost data but as astrophysically informative signals about the spectral energy distribution (SED) shape. This allows the classifier to exploit patterns of non-detection as class-discriminative evidence, paralleling human photometric SED inspection where non-detections in specific bands constrain physical interpretations.

Specifically, \texttt{LightGBM} uses histogram-based, leaf-wise growth for high throughput---well suited to continuous, noisy color features since histogram binning both accelerates training on large synthetic-photometry samples and yields splits less sensitive to noise. \texttt{CatBoost} employs ordered boosting and native categorical feature handling to curb overfitting, along with a symmetric tree structure that improves balanced learning. \texttt{XGBoost} emphasizes explicit regularization and sparse-aware training, which stabilizes splits when non-detections are masked. Additionally, \texttt{XGBoost}'s compatibility with SHAP values enhances model interpretability, which is critical for one of our study's focuses on optimizing filter/feature combinations. The best-performing model from these will serve as our baseline multi-class classifier.


Hyperparameter tuning was conducted using \texttt{Optuna}, which iteratively searched for the best parameter combination over 100 trials to maximize the macro $F_{1}$-score. The macro F1 was defined as:
\begin{equation}
{\rm F1_{macro}} = \frac{1}{K}\sum_{k=1}^K \frac{2 P_k R_k}{P_k + R_k},
\end{equation}
where $P_k$ and $R_k$ are the precision and recall for class \(k\), and \(K\) is the total number of classes. Precision, recall, and accuracy are given by:
\begin{equation}
P_k = \frac{TP_k}{TP_k + FP_k}, \qquad
\end{equation}
\begin{equation}
R_k = \frac{TP_k}{TP_k + FN_k}, \qquad
\end{equation}
\begin{equation}
{\rm Accuracy} = \frac{\sum_k TP_k}{\sum_k (TP_k + FP_k + FN_k)}.
\end{equation}
Here, \(TP_k\), \(FP_k\), and \(FN_k\) represent true positives, false positives, and false negatives for class \(k\), respectively. The $F_{1}$-score is the harmonic mean of precision and recall, offering a balanced measure that penalizes extreme disparities between them. Macro $F_{1}$ thus ensures a uniform evaluation across all classes by equally weighting precision and recall---an essential property for our general-purpose classifier designed to operate within a hybrid KN detection framework, where consistent performance across both rare and common classes is critical.


\begin{figure*}
    \centering
    \includegraphics[width=1\linewidth]{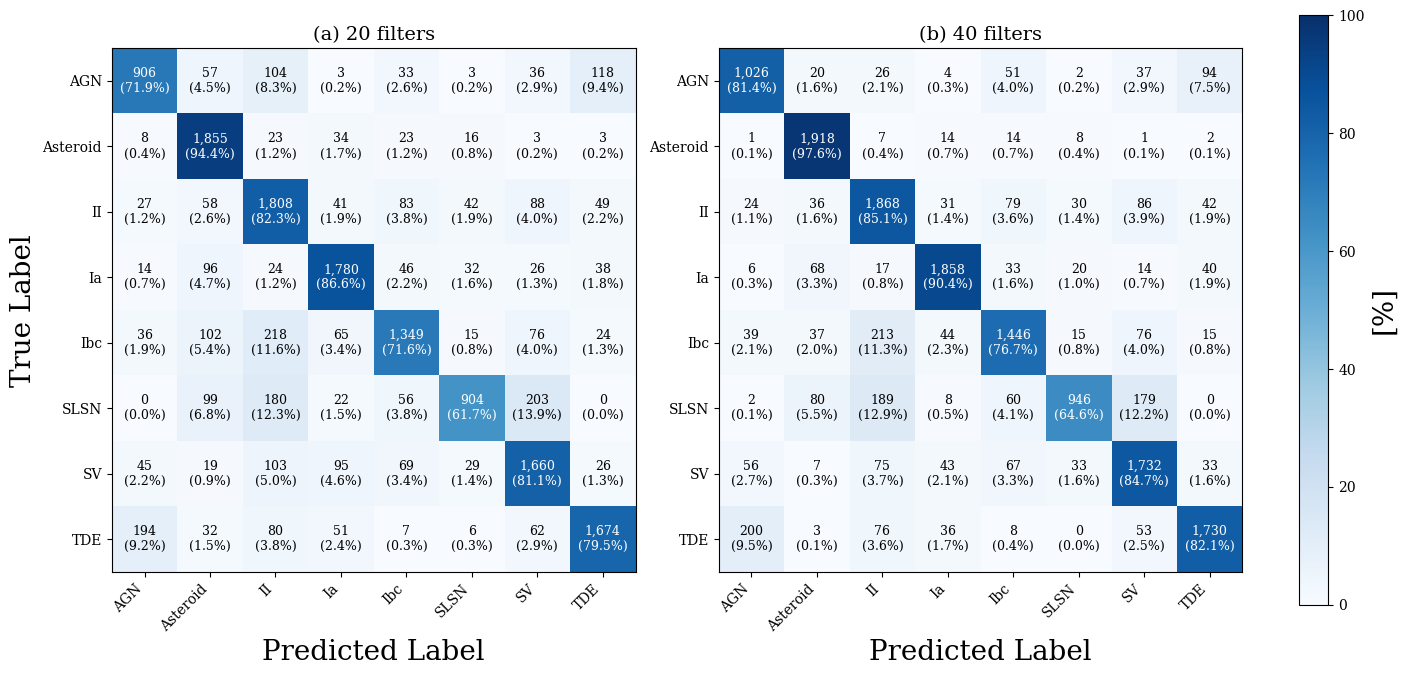}
    \caption{
    Class-normalized confusion matrices of the \texttt{XGBoost} classifier. 
    The left panel shows results using the 7DT 20-filter configuration, while the right panel shows the 40-filter configuration. 
    Each entry indicates the fraction of true class samples classified into the predicted class, normalized by the number of true objects in that class. 
    }
    \label{fig:confusion_matrix}
\end{figure*}

Based on performance (Tables~\ref{tab:result20} and~\ref{tab:result40}), both \texttt{XGBoost} and \texttt{CatBoost} performed strongly; we ultimately selected \texttt{XGBoost} for its robust results and excellent compatibility with SHapley Additive exPlanations (SHAP; \citealt{2017arXiv170507874L}), enabling consistent, locally accurate feature attributions \citep{2023MNRAS.526..404H,2023ApJ...959...44L,2024MNRAS.528..255R} although the $F_{1}$-score of \texttt{CatBoost} is slightly larger than \texttt{XGBoost} in 20-filter set. With this choice, the trained and tuned multi-class classifier achieved a macro $F_{1}$-score of $\sim$0.80 with the 20-filter set and $\sim$0.84 with the 40-filter set. This performance is sufficient for the first-stage filtering in the context of hybrid classification framework, providing a reliable separation of common transients before applying the anomaly detection stage. Figure~\ref{fig:confusion_matrix} shows the confusion matrix for each class.

\subsubsection{Anomaly Classifier}\label{subsubsec:anomalyclassifier}
We use the Isolation Forest (\texttt{iForest}) algorithm \citep{Liu2012} to identify anomalous events such as KNe. It is highly effective in high-dimensional feature spaces and is specifically designed to find outliers without needing to model the complex distribution of all nominal data. This approach has been successfully applied in transient classification \citep{2025MNRAS.543..351C,2025RASTI...4...54G,2024arXiv241018875V}. 


The key principle of the \texttt{iForest} is that anomalies are both rare and distinct, making them easier to isolate than nominal samples \citep{Liu2012}. The algorithm constructs an ensemble of ``Isolation Trees'' by recursively partitioning the feature space using randomly selected features (e.g., color indices) and split values. Data points that require fewer partitions---or equivalently, shorter average path lengths across the ensemble---are identified as anomalies. 


We trained an \texttt{iForest} solely on non-KN samples using default hyperparameters (\texttt{n\_estimators}=100, \texttt{max\_samples}=\texttt{"auto"}, \texttt{contamination}=\texttt{"auto"}) and evaluated it on a hold-out set that included simulated KNe and the observed AT~2017gfo; these KN samples were reserved strictly for evaluation only.
This design exploits the practical advantage of detecting rare classes-even when labeled examples are scarce or model coverage is incomplete-by flagging deviations from the learned ``normal'' distribution of transients. This approach mitigates the risk of bias inherent to supervised models when KN cases are extremely limited in both number and diversity, and when theoretical models may not yet be fully validated.

The resulting anomaly scores ($P_{\rm ano}$) separate KNe from non-KNe, with partial overlap. 
Receiver–operating characteristic (ROC) curves of the true positive rate (TPR) versus the false positive rate (FPR), computed from $P_{\rm ano}$, show competitive KN detection performance (i.e., high TPR at modest FPR; Figure~\ref{fig:iforest}). 
Here, TPR is the fraction of KNe in the hold-out set that are correctly flagged as anomalous for a given $P_{\rm ano}$ threshold, whereas FPR is the fraction of non-KNe incorrectly flagged as anomalous when the decision threshold on $P_{\rm ano}$ is varied:
\[
\mathrm{TPR}=\frac{\mathrm{TP}}{\mathrm{TP}+\mathrm{FN}},\qquad
\mathrm{FPR}=\frac{\mathrm{FP}}{\mathrm{FP}+\mathrm{TN}}.
\]


\subsubsection{Combined Decision Method}\label{subsubsec:combined_decision}

\begin{figure*}
    \centering
    \includegraphics[width=1\linewidth]{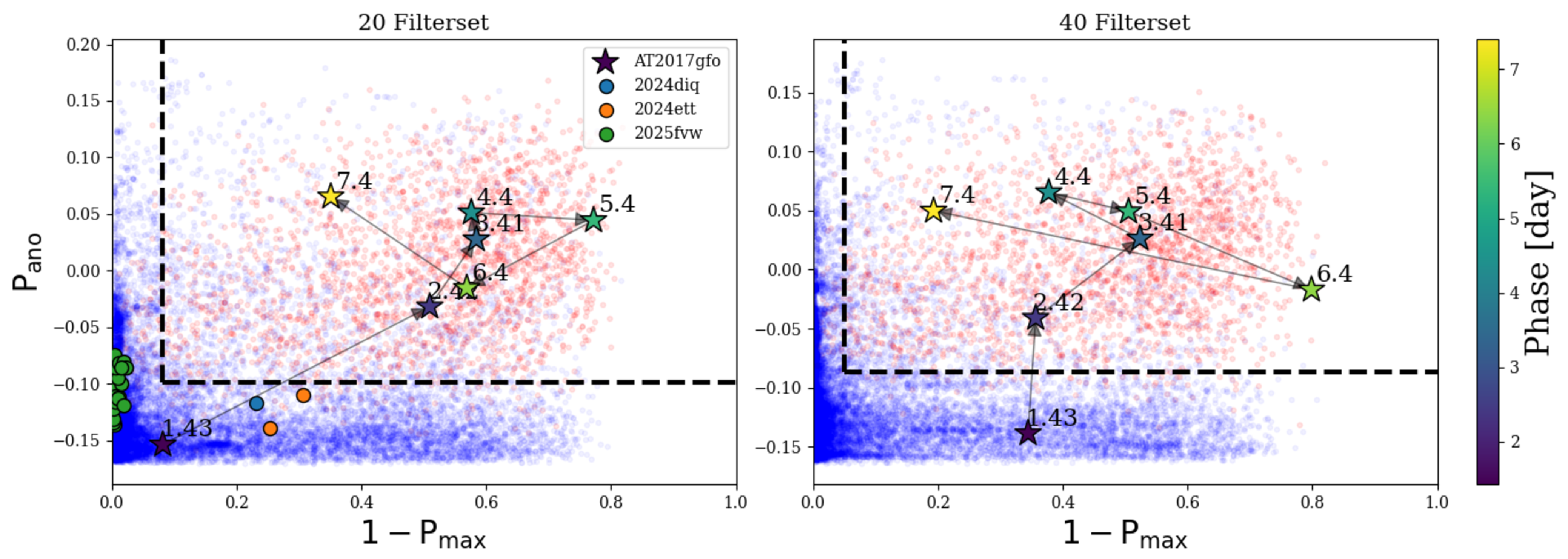}

    \caption{
    Two-dimensional distributions of anomaly score versus classifier uncertainty for the hybrid framework applied to the 20-filter (left) and 40-filter (right) configurations. 
    The horizontal axis is $1-P_{\max}$, where $P_{\max}$ is the highest class probability assigned by the supervised multi-class classifier (\texttt{XGBoost}) to a given input; the vertical axis is the anomaly score $P_{\rm ano}$ returned by the unsupervised detector (\texttt{iForest}). 
    Background points show simulated samples, with blue representing non-KN and red representing KN populations. 
    Colored star markers trace the observed evolution of the KN AT~2017gfo, with color and denoted number next to the marker indicating phase (days), and arrows marking the temporal sequence.
    Blue, orange, and green circles correspond to actual 7DT-observed transients---Type~II SN (SN~2024diq), CV (AT~2024ett), and Type~Ia SN (SN~2025fvw), respectively---as labeled in the legend; these data exist only for the 20-filter configuration (left panel). 
    Dashed lines denote anomaly-classification thresholds determined by Youden's~\,$J$: points below both thresholds are classified as normal, while those exceeding both (logical ``AND'' condition) are classified as anomalies.}
    \label{fig:hybrid}
\end{figure*}

To reduce false positives while retaining high KN recall, we combined the $P_{\mathrm{ano}}$ from \texttt{iForest} (Section~\ref{subsubsec:anomalyclassifier}) with the classifier confidence, $1-P_{\max}$, from the multi-class classifier (Section~\ref{subsubsec:multiclassclassifier}), where $P_{\max}$ is the multi-class classifier's maximum class-probability output and $1-P_{\max}$ quantifies classification uncertainty, thereby defining a two-dimensional decision space (Figure~\ref{fig:hybrid}). Here $P_{\max}$ is the highest class probability among the trained non-KN types; consequently, $1-P_{\max}$ quantifies how weakly the multi-class classifier favors any specific non-KN label for a given input.

\begin{figure}
    \centering
    \includegraphics[width=1\linewidth]{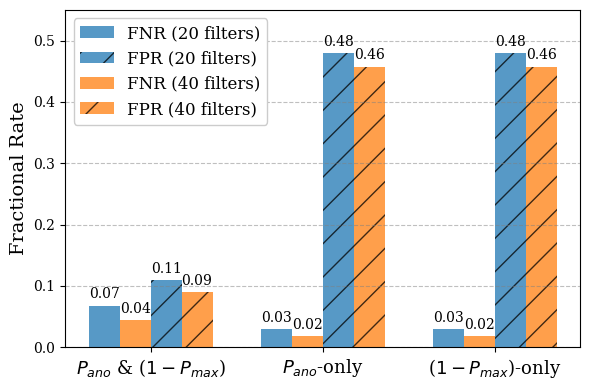}
    \caption{
Error rates of the hybrid framework under different decision rules. 
Solid (non-hatched) bars indicate the FNR (missed KNe), while hatched bars indicate the FPR (non-KNe incorrectly flagged as anomalies). 
Thresholds are determined by Youden's~\,$J$ statistic for three decision cases: 
(1) the joint criterion requiring both the anomaly score and classifier uncertainty ($P_{\rm ano}$ \& $(1-P_{\max})$, ``AND'' gate), 
(2) the anomaly score alone ($P_{\rm ano}$-only), and 
(3) the classifier uncertainty alone ($(1-P_{\max})$-only). 
Results are shown separately for the 20-filter configuration (blue) and the 40-filter configuration (orange), with hatched shading consistently representing FPR values for each.
}
    \label{fig:comp_hybrid}
\end{figure}

To determine the optimal decision thresholds, we adopted Youden's~\,$J$ statistic \citep{1950Ca.....3...32Y} based on the test sample comprising non-KN classes and simulated KNe, defined as

\begin{equation}
    J \equiv \mathrm{Sensitivity} + \mathrm{Specificity} - 1
      = \frac{\mathrm{TP}}{\mathrm{TP+FN}} + \frac{\mathrm{TN}}{\mathrm{TN+FP}} - 1,
\end{equation}

which measures the maximum vertical distance between the ROC curve and the diagonal of random classification. With thresholds, $(1-P_{\max})_{\mathrm{th}},\,P_{\mathrm{ano,th}}=(0.081,-0.099)$ for the 20-filter set and $(0.051,-0.087)$ for the 40-filter set, the framework achieved a KN recall of $0.932$ ($0.955$) and a non-KN false-positive rate of $0.109$ ($0.089$) for the 20 (40) filter configuration (Figure~\ref{fig:comp_hybrid}). The joint rule substantially suppresses false positives while keeping false negatives at acceptable levels.

%
%
%

\section{Results} \label{sec:results}

\subsection{Multi-class Classification Performance}\label{subsec:multiclassclassifier}

\begin{table*}
    \centering
    \caption{Performance comparison using the 20 7DT filter set and their subsets. 
    The complete hyperparameter sets for all models are provided in Appendix~\ref{app:hyperparameter}.}
    \label{tab:result20}
    \begin{tabular}{lcccccc}    
        \toprule
        Test Name & $N_{\mathrm{filter}}$ & $N_{\mathrm{feature}}$ & $F_{1}$-score & Accuracy & Precision & Recall \\
        \midrule
        \multicolumn{7}{c}{\textbf{Full 20-filter Models}} \\
        \midrule
        \textbf{XGBoost}            & 20 & 190 & 0.797 & 0.801 & 0.786 & 0.789 \\
        CatBoost           & 20 & 190 & 0.801 & 0.806 & 0.791 & 0.795 \\
        LightGBM           & 20 & 190 & 0.793 & 0.799 & 0.782 & 0.785 \\
        \midrule
        \multicolumn{7}{c}{\textbf{Feature Subsets from \texttt{XGBoost}}} \\
        \midrule
        Top 10               & 10 & 45 & 0.705 & 0.705 & 0.688 & 0.690 \\
        Bottom 10            & 10 & 45 & 0.637 & 0.639 & 0.635 & 0.634 \\
        10\%  &  9 &  5  & 0.613 & 0.607 & 0.602 & 0.603 \\
        25\%  & 13 & 16  & 0.742 & 0.740 & 0.729 & 0.731 \\
        50\%  & 19 & 45  & 0.792 & 0.799 & 0.782 & 0.785 \\
        75\%  & 20 & 92  & 0.799 & 0.804 & 0.787 & 0.791 \\
        90\%  & 20 & 134 & 0.802 & 0.807 & 0.791 & 0.795 \\
        \midrule
        \multicolumn{7}{c}{\textbf{Top20 7DT + Rubin Broad-band}} \\
        \midrule
        7DT + Rubin $u$ & 21 & 210 & 0.814 & 0.817 & 0.804 & 0.806 \\
        7DT + Rubin $g$ & 21 & 210 & 0.798 & 0.803 & 0.787 & 0.790 \\
        7DT + Rubin $r$ & 21 & 210 & 0.799 & 0.804 & 0.788 & 0.791 \\
        7DT + Rubin $i$ & 21 & 210 & 0.805 & 0.810 & 0.792 & 0.797 \\
        7DT + Rubin $z$ & 21 & 210 & 0.809 & 0.813 & 0.801 & 0.803 \\
        7DT + Rubin $y$ & 21 & 210 & 0.810 & 0.814 & 0.802 & 0.804 \\
        \bottomrule
    \end{tabular}
\end{table*}

\begin{table*}
    \centering
    \caption{Performance comparison using 40 7DT filter set and their subsets. 
    The complete hyperparameter sets for all models are provided in Appendix~\ref{app:hyperparameter}.}
    \label{tab:result40}
    \begin{tabular}{lcccccc}
        \toprule
        Test Name & $N_{\mathrm{filter}}$ & $N_{\mathrm{feature}}$ & $F_{1}$-score & Accuracy & Precision & Recall \\
        \midrule
        \multicolumn{7}{c}{\textbf{Full Filter Models}} \\
        \midrule
        \textbf{XGBoost}                 & 40 & 780 & 0.836 & 0.839 & 0.828 & 0.830 \\
        CatBoost                & 40 & 780 & 0.834 & 0.839 & 0.825 & 0.828 \\
        LightGBM                & 40 & 780 & 0.832 & 0.836 & 0.824 & 0.827 \\
        \midrule
        \multicolumn{7}{c}{\textbf{Feature Subsets from \texttt{XGBoost}}} \\
        \midrule
        Top 20                   & 20 & 190 & 0.796 & 0.796 & 0.784 & 0.786 \\
        Bottom 20                & 20 & 190 & 0.716 & 0.716 & 0.708 & 0.708 \\
        10\%                    & 20 &  16 & 0.770 & 0.766 & 0.759 & 0.760 \\
        20\%                    & 27 &  41 & 0.808 & 0.810 & 0.801 & 0.801 \\
        25\%                    & 28 &  57 & 0.818 & 0.820 & 0.810 & 0.811 \\
        30\%                    & 30 &  75 & 0.826 & 0.831 & 0.816 & 0.819 \\
        40\%                    & 36 & 116 & 0.839 & 0.844 & 0.832 & 0.833 \\
        50\%                    & 40 & 166 & 0.838 & 0.843 & 0.830 & 0.831 \\
        75\%                    & 40 & 347 & 0.837 & 0.840 & 0.829 & 0.831 \\
        90\%                    & 40 & 525 & 0.837 & 0.840 & 0.828 & 0.831 \\
        \midrule
        \multicolumn{7}{c}{\textbf{7DT + Rubin Broad-band}} \\
        \midrule
        7DT+Rubin $u$           & 41 & 820 & 0.840 & 0.845 & 0.832 & 0.834 \\
        7DT+Rubin $g$           & 41 & 820 & 0.838 & 0.842 & 0.829 & 0.831 \\
        7DT+Rubin $r$           & 41 & 820 & 0.834 & 0.838 & 0.825 & 0.827 \\
        7DT+Rubin $i$           & 41 & 820 & 0.838 & 0.842 & 0.829 & 0.832 \\
        7DT+Rubin $z$           & 41 & 820 & 0.835 & 0.838 & 0.829 & 0.831 \\
        7DT+Rubin $y$           & 41 & 820 & 0.843 & 0.847 & 0.837 & 0.839 \\
        \midrule
        \multicolumn{7}{c}{\textbf{7DT Top20 + Rubin Broad-band$^{\dagger}$}} \\
        \midrule
        7DT+Rubin $u^{\dagger}$ & 21 & 210 & 0.811 & 0.813 & 0.800 & 0.801 \\
        7DT+Rubin $g^{\dagger}$ & 21 & 210 & 0.799 & 0.802 & 0.787 & 0.790 \\
        7DT+Rubin $r^{\dagger}$ & 21 & 210 & 0.796 & 0.797 & 0.786 & 0.787 \\
        7DT+Rubin $i^{\dagger}$ & 21 & 210 & 0.817 & 0.821 & 0.810 & 0.812 \\
        7DT+Rubin $z^{\dagger}$ & 21 & 210 & 0.819 & 0.823 & 0.813 & 0.815 \\
        7DT+Rubin $y^{\dagger}$ & 21 & 210 & 0.823 & 0.827 & 0.820 & 0.819 \\
        \bottomrule
    \end{tabular}
    \vspace{0.3em}
    \begin{flushleft}
        \footnotesize{$^{\dagger}$Subset using SHAP Top20 7DT filters with a single Rubin broad-band.}
    \end{flushleft}
\end{table*}

We first evaluate the overall performance of the supervised multi-class classifier trained on synthetic 7DT photometry across eight transient classes. Using macro $F_{1}$ score, the \texttt{XGBoost} model achieves $F_1 \approx 0.80$ with the 20-filter configuration and $F_1 \approx 0.82$ with the 40-filter configuration (Table~\,\ref{tab:result20} and \ref{tab:result40}). 
As shown in Figure~\ref{fig:confusion_matrix}, Type~Ia SNe are classified robustly, with $F_{1}\approx 0.87$ and $0.90$ for the 20- and 40-filter sets, respectively. Type~Ibc SNe and Type~II SNe show bidirectional confusion at the $\sim$11\% level, and SLSNe are more frequently mistaken for Type~II SNe or SV. Overall, the 40-filter configuration improves class-wise $F_{1}$ by a few up to $\sim$10~\,\% points relative to the 20-filter setup, with the largest gain observed for AGN (by $\rm \sim10\%$).

\subsection{Feature and Filter Importance Analysis from SHAP Values} \label{subsec:importance}
We analyzed SHAP-based feature importance \citep{2017arXiv170507874L,2019arXiv190504610L} to (i) test whether a reduced feature set can retain classification performance and thereby identify optimized filter combinations, and (ii) examine whether the model's attribution of importance to spectral features is consistent with physical intuition (e.g., broad SED structures and diagnostic spectral lines).


\begin{figure*}
    \centering
    \includegraphics[width=0.8\linewidth]{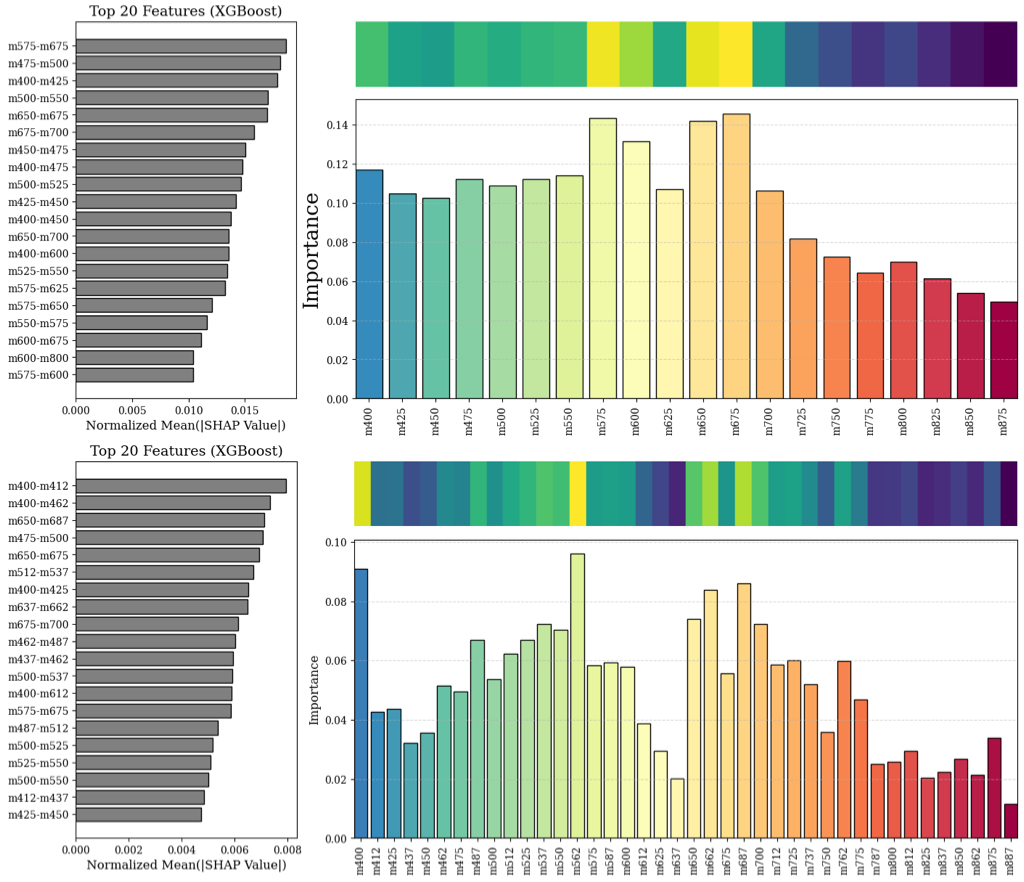}
    \caption{
    Feature-to-filter importance conversion for the \texttt{XGBoost} classifier. 
    Left panels show the top 20 color features ranked by their mean absolute SHAP values, while right panels present the corresponding filter-level importance obtained by decomposing the feature contributions into their constituent filters. 
    The horizontal bar plots (left) highlight the most discriminative color combinations, and the bar charts (right) illustrate the relative weight of each filter in the classification. 
    Color strips above each bar chart visualize the filter positions in wavelength space, with brighter regions corresponding to higher importance. 
    The top row corresponds to the 20-filter configuration and the bottom row to the 40-filter configuration.
    }
    \label{fig:feature_importance}
\end{figure*}

To obtain filter-level importance from color-feature SHAP scores, we decomposed each color feature (e.g., $m512-m537$) into its two constituent bands and assigned the feature's SHAP importance as a weight to both bands; summing these weights over all features yields a per-filter importance score as shown in Figure~\ref{fig:feature_importance}. Using these filter scores, to validate the connection between feature importance and model performance, we further compared classification results obtained using only the top- and bottom-ranked filters according to SHAP importance. For the 20-filter set, the top-10 subset outperformed the bottom-10 subset by $\sim$5\%, while for the 40-filter set the top-20 subset exceeded the bottom-20 subset by $\sim$8\%, confirming that the ranked importance correlates with predictive power.

\begin{figure*}
    \centering
    \includegraphics[width=1\linewidth]{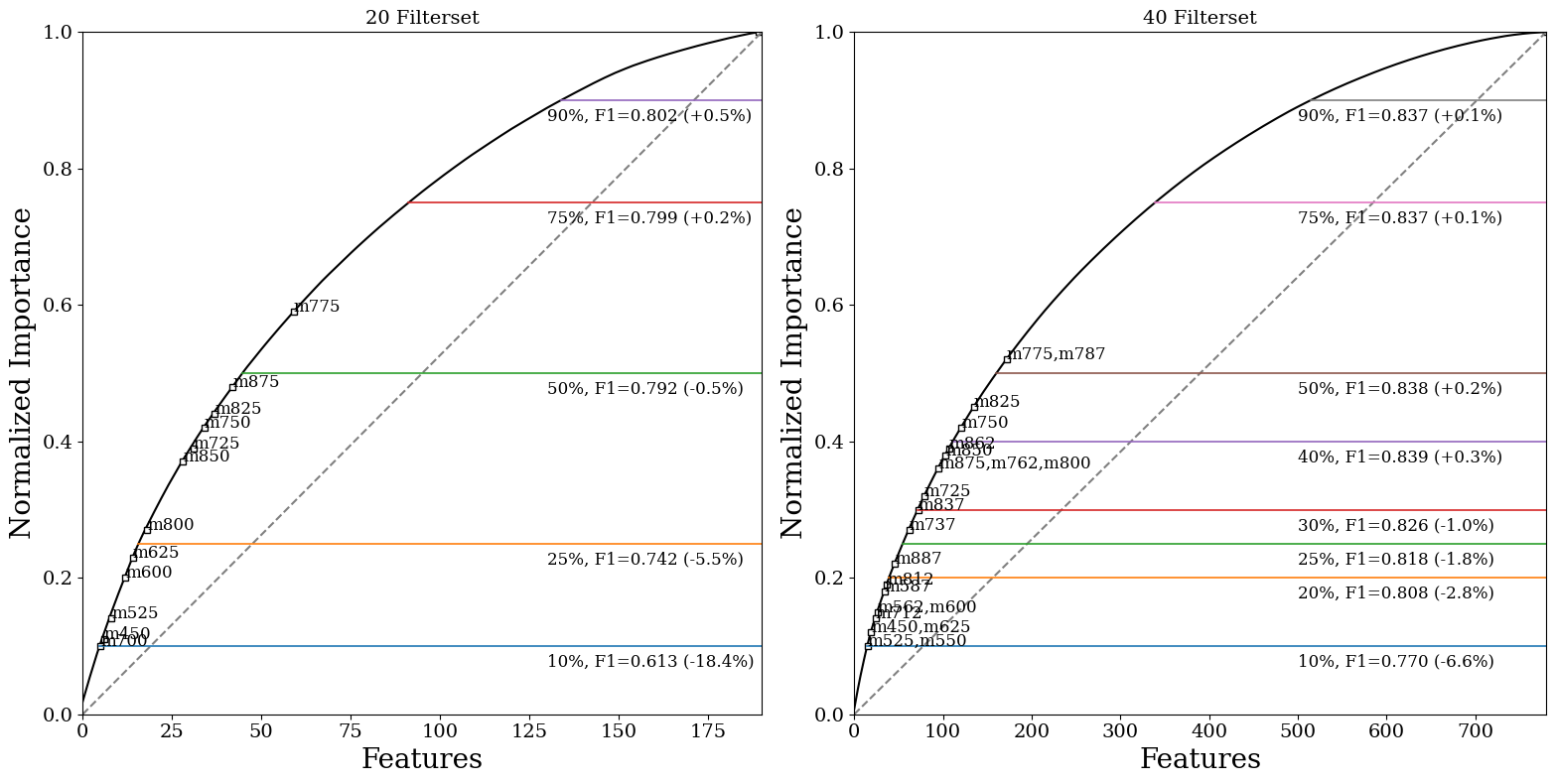}
    \caption{
    Cumulative distribution of feature importance derived from SHAP values for the \texttt{XGBoost} classifier. 
    The left panel shows the 20-filter configuration and the right panel shows the 40-filter configuration. 
    Black curves indicate the normalized cumulative importance of all color features, with dashed gray lines marking uniform importance for reference. 
    Horizontal colored lines denote the performance of models retrained using only the top $N$\% of features, with the corresponding $F_{1}$ score (see Table~\,\ref{tab:result20} and \ref{tab:result40} for each 20 and 40 filter set) and relative change compared to the full model annotated. 
    Filter labels along the cumulative curve mark the first instance at which those filters are entirely excluded from the selected feature set as $N$\% decreases. 
    }
    \label{fig:feature_cumulative_distribution}
\end{figure*}

Following the cumulative distribution of feature importance, we trained models using only the top fraction of filters and evaluated their performance (Figure~\ref{fig:feature_cumulative_distribution}). Performance remained nearly unchanged-and in some cases slightly exceeded the baseline when intermediate subsets were used-before degrading as the retained fraction became too small. In the 20-filter configuration, degradation became apparent only when restricting to the top 50\% of filters; at that threshold the excluded filter is $m775$. In the 40-filter configuration, performance began to drop when restricting to fewer than the top 40\% of filters; at that cutoff the excluded filters are $m750$, $m775$, $m787$, and $m825$.



%

\subsection{KN Anomaly Detection Performance}\label{subsec:anomalyclassifier}



%



\subsection{Validation on Test Samples}

To assess the robustness and practical applicability of our hybrid framework, we validated its performance on both simulated and real data sets. This includes a broad grid of synthetic KN models spanning diverse ejecta parameters (Section~\,\ref{subsubsec:simkn}), the synthesized data from observed KN AT~2017gfo (Section~\,\ref{subsubsec:engrave}), and a set of non-KN transients from 7DT observations (Section~\,\ref{subsec:7dt_test}). 

\subsubsection{Broad Grid of Simulated KNe}

\begin{figure*}
    \centering
    \includegraphics[width=1\linewidth]{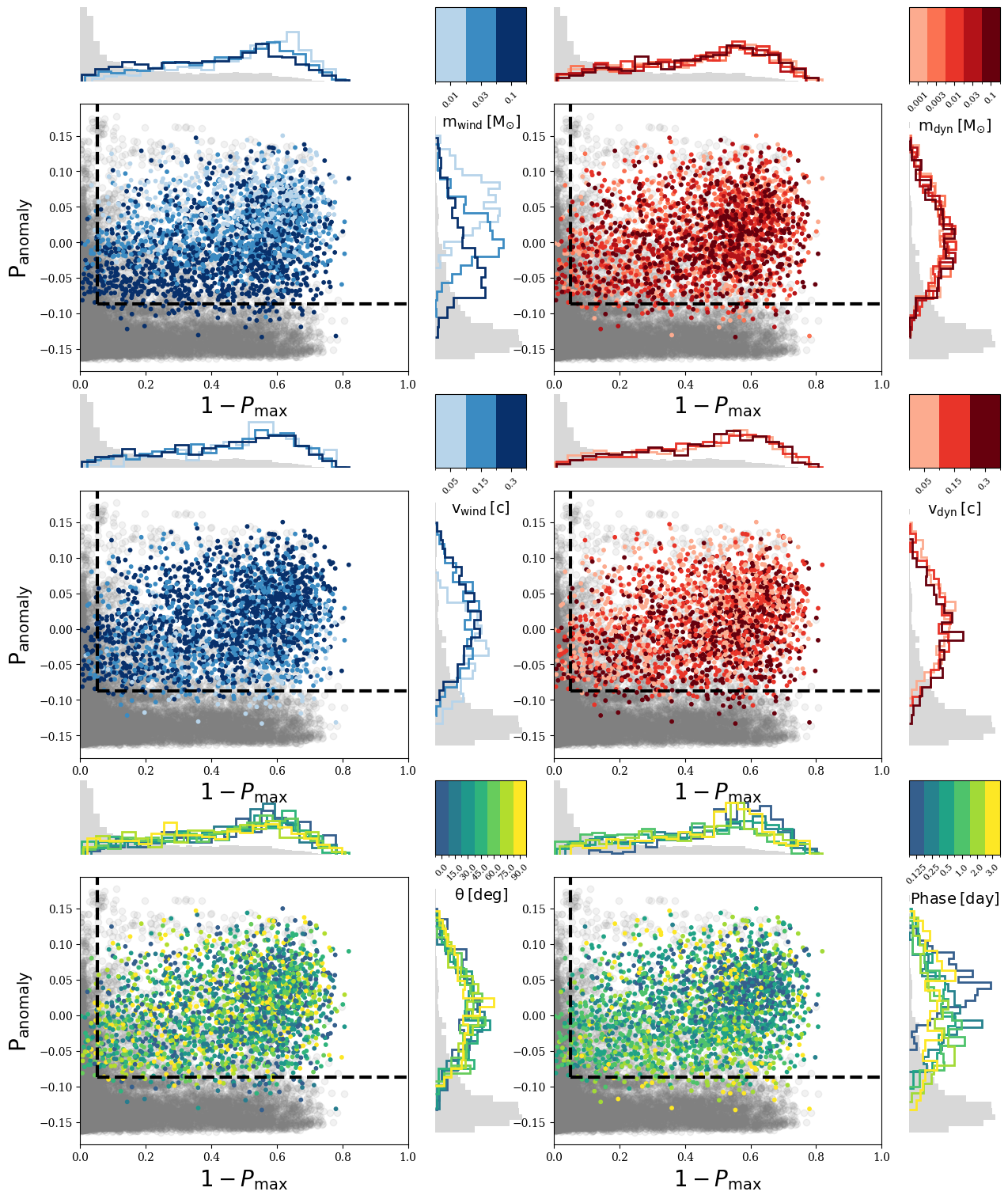}
    \caption{
    Hybrid framework decision space for the 40-filter configuration. 
    Non-KN test samples are shown in gray, while simulated KNe are overplotted and color-coded by ejecta parameters from the synthetic grid: wind ejecta mass ($m_{\rm wind}$), wind velocity ($v_{\rm wind}$), dynamical ejecta mass ($m_{\rm dyn}$), dynamical ejecta velocity ($v_{\rm dyn}$), viewing angle ($\theta$), and phase (days since merger). 
    The dashed lines indicate thresholds in $P_{ano}$ and $1-P_{\max}$ determined by maximizing Youden's~\,$J$ statistic. 
    Marginal histograms highlight the distributions of KNe in each parameter subset relative to the background of non-KN transients.
    }
    \label{fig:hybrid_kn_sim40}
\end{figure*}

Applying the hybrid scoring to the synthetic KN grid shows that wind ejecta properties impact the anomaly score (Figure~\,\ref{fig:hybrid_kn_sim20} and \ref{fig:hybrid_kn_sim40}): higher wind mass and lower wind velocity tend to reduce the $P_{ano}$, while the classifier confidence $1-P_{\max}$ remains comparatively stable. Not only that, the earliest phase (0.125~\,days) shows the higher $P_{ano}$ scores than other phases. Later phases show no significant dependence on the $P_{ano}$. By contrast, viewing angle, and dynamical ejecta parameters have weaker influence on both scores.

\subsubsection{AT~2017gfo KN}

\begin{deluxetable*}{ccccccccc}
\tablecaption{Classification results with 20- and 40-filter sets \label{tab:classification}}
\tablewidth{0pt}
\tablehead{
  \colhead{\begin{tabular}{c}Phase\\ {(day)}\end{tabular}} &
  \colhead{$P_{20,\max}$} &
  \colhead{Class$_{20}$} &
  \colhead{$P_{20,\mathrm{ano}}$} &
  \colhead{Flag$_{\mathrm{ano},20}$} &
  \colhead{$P_{40,\max}$} &
  \colhead{Class$_{40}$} &
  \colhead{$P_{40,\mathrm{ano}}$} &
  \colhead{Flag$_{\mathrm{ano},40}$}
}
\startdata
1.43 & 0.918 & Asteroid & $-0.154$ &        & 0.655 & Asteroid & $-0.139$ &        \\
2.42 & 0.490 & AGN      & $-0.032$ & $\checkmark$ & 0.643 & AGN      & $-0.041$ & $\checkmark$ \\
3.41 & 0.416 & SV       &  0.027   & $\checkmark$ & 0.476 & SV       &  0.026   & $\checkmark$ \\
4.40 & 0.424 & SV       &  0.051   & $\checkmark$ & 0.622 & SV       &  0.065   & $\checkmark$ \\
5.40 & 0.228 & II       &  0.044   & $\checkmark$ & 0.494 & AGN      &  0.049   & $\checkmark$ \\
6.40 & 0.431 & SLSN     & $-0.016$ & $\checkmark$ & 0.201 & AGN      & $-0.017$ & $\checkmark$ \\
7.40 & 0.649 & SLSN     &  0.065   & $\checkmark$ & 0.807 & SLSN     &  0.050   & $\checkmark$ \\
\enddata
\tablecomments{%
  Columns list maximum class probability and anomaly score (``ano'') for
  the 20- and 40-filter sets. Flags use a checkmark ($\checkmark$) to indicate anomalous cases.
}
\end{deluxetable*}

For AT~2017gfo, the earliest spectrum ($\sim$1.43~day) lies near the overlap region with non-KNe in the space of two metrics and can be misclassified as by the multi-class classifier alone as an asteroid, whereas all later epochs are clearly flagged as anomalous by the hybrid framework (Figure~\ref{fig:hybrid}). 

\subsubsection{Test with 7DT observations}
We applied the framework to three transients (SN~2025fvw; Type~Ia SN, SN~2024diq; Type~II SN, and AT~2024ett; CV observed by 7DT, described in Section \ref{subsec:7dt}) obtained with the 20-filter set configuration. All were correctly retained as non-anomalous. The multi-class classifier labels were consistent with their known types across epochs for SN\,2025fvw, but SN\,2024diq and AT\,2024ett were misclassified as TDE and Type Ia SN respectively.

\subsection{Synergy with LSST photometry} \label{subsec:rubin}

We computed synthetic LSST ($ugrizy$) photometry using \texttt{speclite} to enable joint 7DT–Rubin classification tests. For each input spectrum, fluxes were integrated over the LSST filter curves and SNRs estimated using the nominal 5$\sigma$ depths \citep{2022ApJS..258....1B} ($u=23.9,\, g=25.0,\, r=24.7,\, i=24.0,\, z=23.3,\, y=22.1$). Because some archival spectra lacked full $u$/$y$-band coverage, we linearly extrapolated them to cover the missing wavelength regions. To simulate real-time follow-up after LSST alerts, we generated cross-system color features by combining single-band LSST photometry with 7DT medium-band data, assuming negligible SED evolution between the two epochs.

The largest performance gains were achieved by adding the LSST $u$ band to the 20-filter set and the $y$ band to the 40-filter set, though these were modest ($\sim$1\%) and statistically insignificant. When combining the SHAP-ranked top 20 filters with one LSST band, performance stayed within $\sim$4\% of the 40-filter baseline, with the $y$ band offering the largest boost (2.7\%).

%
%

\section{Discussion} \label{sec:discussion}
Our primary objective is to reliably recover KNe as anomalies within a heterogeneous transient stream. To this end, we adopt a hybrid framework that applies an unsupervised anomaly detector (\texttt{iForest}) on single-epoch, medium-band 7DT SEDs, guided by a supervised multi-class classifier (\texttt{XGBoost}) that models the dominant non-KN classes. The overall macro performance of the multi-class classifier reaches $F_{1}\!\approx\!0.80$--$0.82$ for the 20- and 40-filter configurations (Tables~\ref{tab:result20} and \ref{tab:result40}), and the hybrid framework effectively classifies KN test samples---without using KNe in training---as anomalies distinct from non-KN classes, yielding a low contaminant fraction. This indicates that a single low-resolution SED snapshot is sufficiently informative to separate diverse transient populations. 

Consistent with the findings of \citet{2025arXiv250722106F}, who demonstrated that even low-resolution SEDs ($R\sim$30) retain sufficient diagnostic power to capture SNe spectral features and overall continuum shapes, our results confirm that realistic sensitivity modeling in 7DT yields robust performance across not only SNe but also other types of transients. Thus, our framework not only substantiates earlier claims of the utility of low-resolution SEDs for classification but also extends them to a realistic survey setting with instrument-specific throughput and detection limits. Below, we interpret the model, quantify operational trade-offs, and discuss limitations and implications for KN searches.

\subsection{Physical Interpretation of the Learned Features} \label{subsec:interpretation}




\begin{figure*}
    \centering
    \includegraphics[width=1\linewidth]{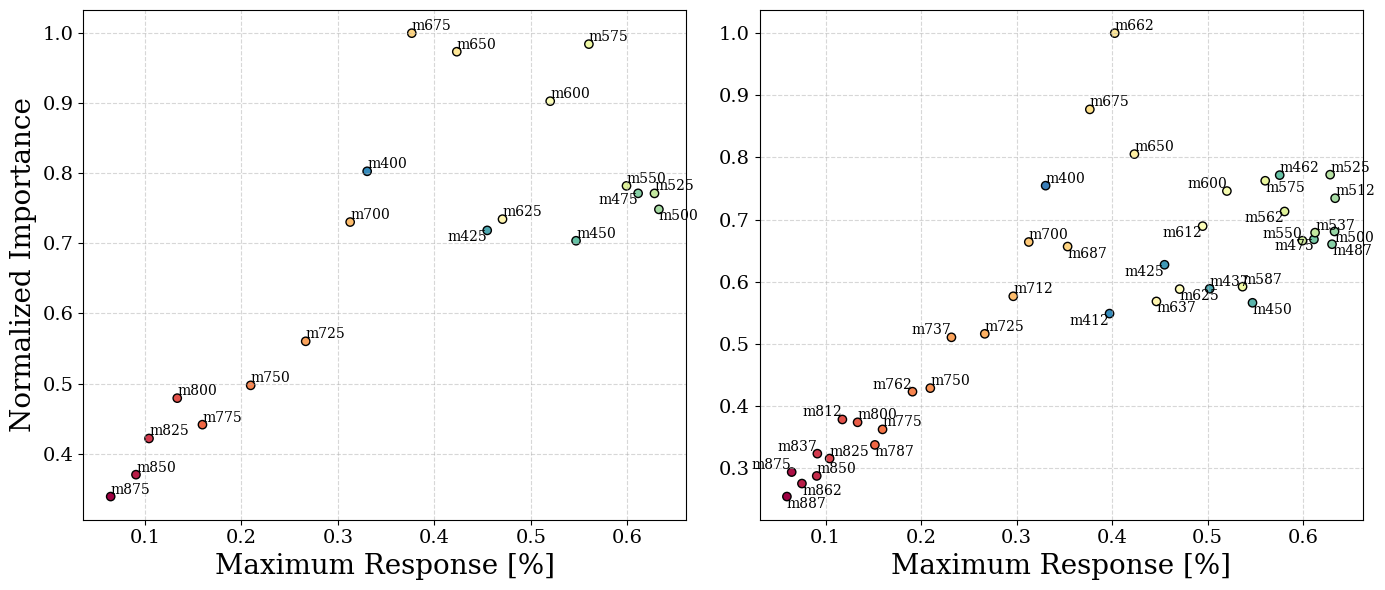}

    \caption{
    Correlation between the normalized feature importance (from SHAP analysis of the \texttt{XGBoost} classifier) and the maximum filter response value for the 7DT medium-band filters. 
    The left and right panels correspond to the 20- and 40-filter configurations, respectively. 
    The $x$-axis shows the maximum throughput of each filter response curve (as defined in Figure~2), while the $y$-axis shows the normalized importance of each filter in the classification task. 
    }
    
    \label{fig:imp_vs_rsp}
\end{figure*}

The analysis of filter importance derived from SHAP in the multi-class classifier highlights their complementary roles in classification. 
To verify whether these qualitative expectations translate into model performance, we retrained \texttt{XGBoost} using the top N\% of filters based on the SHAP value. As a result, excluding low-importance red-end bands (e.g., $m750$, $m775$, $m787$, $m825$) does not degrade, and in some cases even improves, performance (Tables~\ref{tab:result20} and \ref{tab:result40}). Furthermore, restricting the model to the top 40\% of features (36 among 40 filters) as shown in Figure~\ref{fig:feature_cumulative_distribution} maintains macro $F_{1}$ of the full model.

A SHAP-based ranking provides the underlying interpretation: constraining the classifier to the top $\sim$10--25\% of features in 40 filter set configuration generally preserves-and occasionally improves-performance (Tables~\ref{tab:result20} and \ref{tab:result40}). 
Informative bands cluster around strong spectral features, especially H$\alpha$ ($m650$--$m675$), while filters redward of $\sim$725\,nm contribute less on average owing to lower instrumental sensitivity and larger photometric scatter. 
The correlation between importance and instrumental throughput (Figure~\ref{fig:imp_vs_rsp}) indicates that astrophysical leverage and detector response jointly determine utility; consequently, compact, importance-guided subsets suffice for efficient classification and triage.

We investigated why longer wavelength filters have less important in this classification context.
As shown in Figure~\ref{fig:imp_vs_rsp}, filters with higher sensitivity contribute more strongly to classification, since their lower scatter reduces color-related confusion during model training. 
Filters with a maximum throughput below 30\% ($>$700~\,nm) show both low and linearly correlated normalized importance, implying that instrumental sensitivity is a primary factor governing model performance.

\begin{figure*}
    \centering
    \includegraphics[width=1\linewidth]{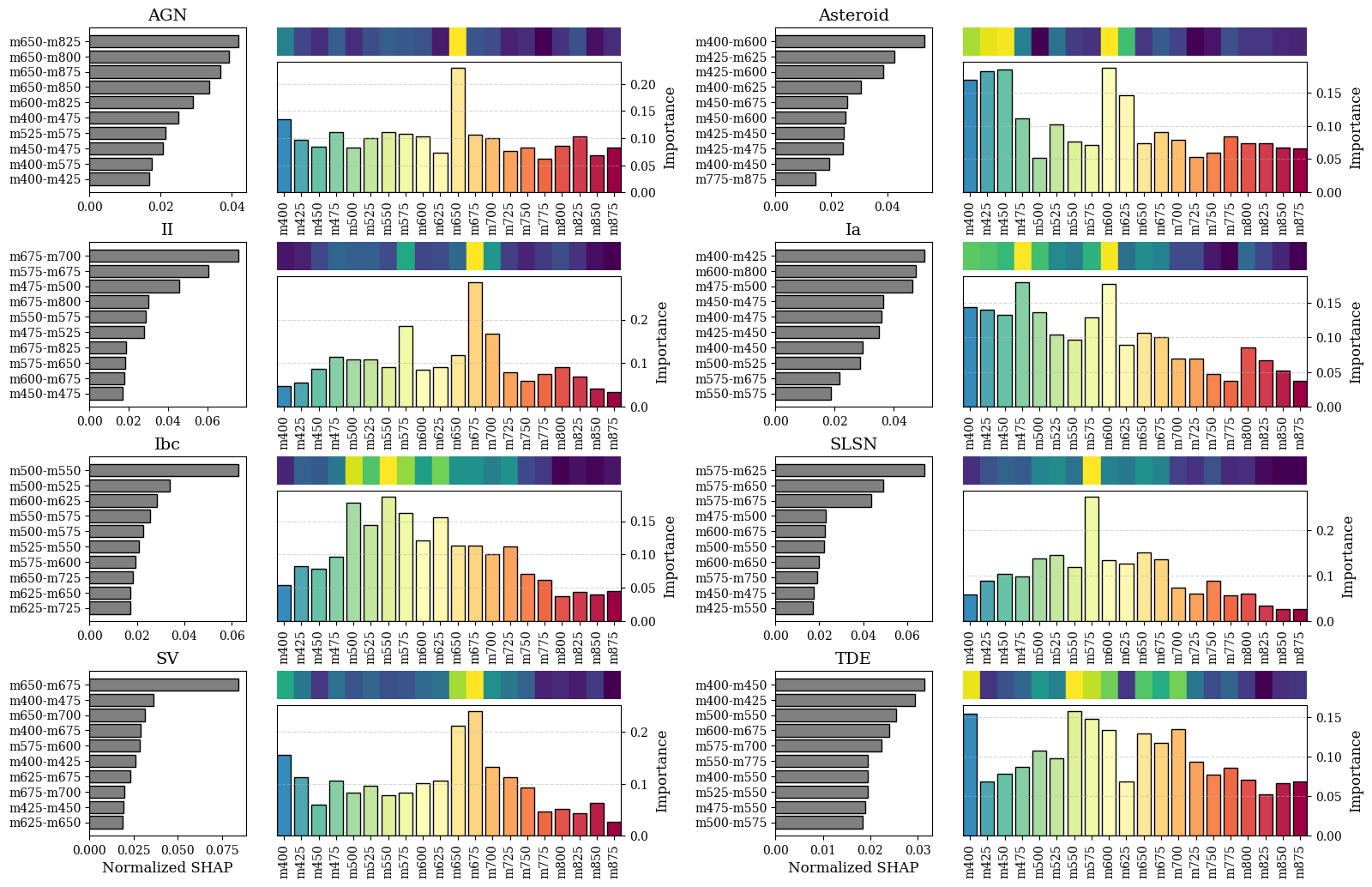}
    \caption{
    Class-decomposed SHAP-based feature and filter importance for the \texttt{XGBoost} classifier for 20 filter set. 
    For each transient class (AGN, Asteroid, Type~II, Ia, Ibc, SLSN, SV, TDE), the left panels show the top-ranked color features (e.g., $m600{-}m625$) according to their normalized mean absolute SHAP values, while the right panels present the aggregated filter-level importance obtained by decomposing each color feature into its constituent filters. 
    The horizontal bar plots highlight which color combinations drive the classification of each class, whereas the vertical bar charts reveal the relative contributions of individual filters. 
    Color strips above each bar chart indicate the filter positions in wavelength space, with brighter regions corresponding to higher importance. 
    }
    \label{fig:class_decomposed20}
\end{figure*}

\begin{figure*}
    \centering
    \includegraphics[width=1\linewidth]{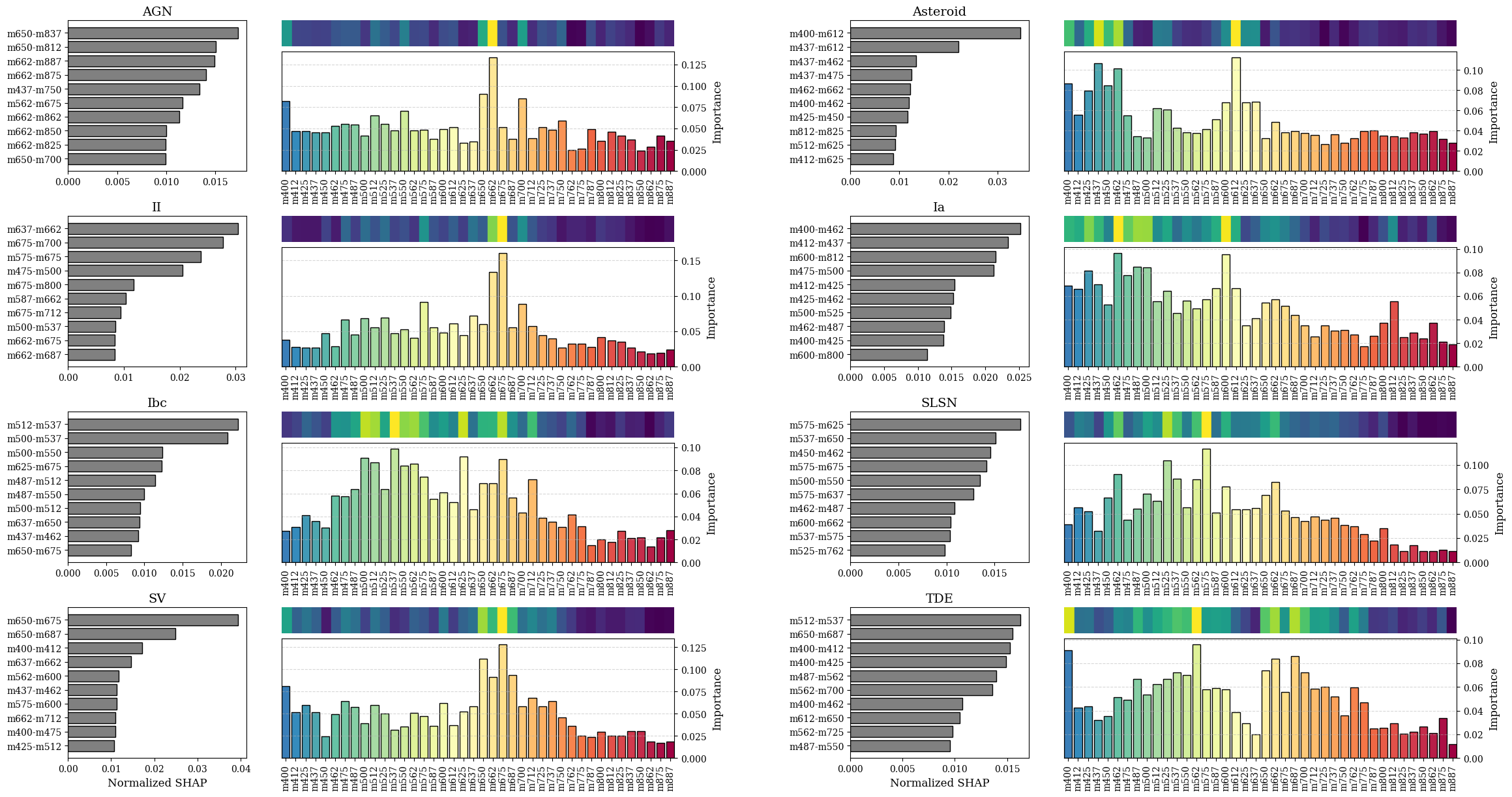}
    \caption{
    Class-decomposed SHAP-based feature and filter importance for the \texttt{XGBoost} classifier for 40 filter set. 
    For each transient class (AGN, Asteroid, Type~II, Ia, Ibc, SLSN, SV, TDE), the left panels show the top-ranked color features (e.g., $m600{-}m625$) according to their normalized mean absolute SHAP values, while the right panels present the aggregated filter-level importance obtained by decomposing each color feature into its constituent filters. 
    The horizontal bar plots highlight which color combinations drive the classification of each class, whereas the vertical bar charts reveal the relative contributions of individual filters. 
    Color strips above each bar chart indicate the filter positions in wavelength space, with brighter regions corresponding to higher importance. 
    }
    \label{fig:class_decomposed40}
\end{figure*}

Class-decomposed importance maps illustrate how filter-level trends manifest across transient categories (Figures~\ref{fig:class_decomposed20} and \ref{fig:class_decomposed40}). The patterns are consistent with known spectroscopic diagnostics:
\begin{itemize}
  \item \textbf{Type Ia SN}: elevated weight around the Si\,\textsc{ii}\,$\lambda6355$ absorption trough (near $\sim$6150\,\AA; e.g., $m600$--$m625$) and in the blue ($m400$--$m462$).
  \item \textbf{Type Ibc SN}: reliance on broad continuum slopes and weaker IME blends (e.g., $m500$--$m550$); degeneracy with Type~II persists at single-epoch, low spectral resolution.
  \item \textbf{Type II SN}: emphasis near H$\alpha$ and adjacent complexes (e.g., $m637$--$m662$, $m675$--$m700$), as expected for hydrogen-rich spectra.
  \item \textbf{SLSN}: preference for blue bands (e.g., $m450$--$m462$), reflecting hot continua and sensitivity to O\,\textsc{ii} complexes in the $\sim$3500--4500\,\AA\ range.
  \item \textbf{TDE}: elevated importance in the blue optical ($m400$--$m512$) and around H$\alpha$, consistent with hot continua plus Balmer and He\,\textsc{ii}\,$\lambda4686$ emission in nuclear flares.
  \item \textbf{AGN}: peaks around the H$\alpha$ neighborhood and continuum-slope tracers; in medium bands, blended narrow-line complexes (H$\alpha$+[N\,\textsc{ii}],[S\,\textsc{ii}]) shape the SED more than line resolution per se.
  \item \textbf{Asteroid}: G2V-like reflected-solar SED with a blue-ward decline; long-baseline colors (e.g., $m400$--$m612$) provide strong leverage.
  \item \textbf{Stellar variable (SV)}: importance spread across $m500$--$m675$, consistent with class heterogeneity and mixed Balmer/continuum regions.
\end{itemize}
Taken together, the model prioritizes bands that bracket diagnostic features (e.g., H$\alpha$, Si\,\textsc{ii}) or anchor continuum curvature; residual degeneracies are expected at $R\sim20$--35 with single-epoch SEDs.

Individual filters may contribute through (i) effective sensitivity-set both by instrumental throughput and by the source brightness at that wavelength (i.e., lower scatter), (ii) placement on or near distinctive spectral features or continuum ``anchor'' regions that bracket such features, and (iii) inter-class separability when combined with other bands. In practice, these factors interact, and overlaps in line/continuum regions or shared SED shapes introduce degeneracies; thus, filter (or feature) importance typically reflects a mixture of sensitivity, spectral leverage, and discriminative power rather than a single cause; however, the dominant drivers align with intuition---heightened sensitivity, human-used diagnostics such as H$\alpha$, and long-baseline colors that encode global continuum shape---indicating that the model has learned in an interpretable, physically meaningful manner.

\subsection{KN Detection Performance and Validation}\label{subsec:kn_det}


The $P_{ano}$ from \texttt{iForest} and the uncertainty proxy $1-P_{\max}$ from the Multi-class classifier probe different aspects of model failure and are not expected to correlate linearly. \texttt{iForest} recursively partitions feature space and isolates outliers in short path lengths \citep{Liu2012}, while $1-P_{\max}$ measures ambiguity among trained classes from \texttt{XGBoost} \citep{2016arXiv160302754C}. 
A transient may therefore be structurally rare yet confidently mapped to a known class, or be ambiguous without being an outlier. 
Jointly thresholding the 2D score space via Youden's~\,$J$ reduces non-KN false positives while maintaining high KN recall, yielding tangible gains in follow-up efficiency. 
Notably, if one relies solely on $P_{ano}$ from \texttt{iForest}, the FPR reaches nearly 50\%, leading to substantial contamination of KN candidates as shown in Figure~\,\ref{fig:comp_hybrid}. By contrast, incorporating $1-P_{\max}$ in the hybrid framework reduces the FPR to $\sim$10\% while preserving nearly the same KN recall, underscoring the critical advantage of combining these complementary diagnostics.
We tested the hybrid framework using simulated KNe spanning a broad parameter space and real observations, including AT~2017gfo and 7DT transients. 

\subsubsection{AT~2017gfo KN} 
For AT~2017gfo, most epochs in both filter sets are successfully classified as anomalous, falling within the Youden's~\,$J$ thresholds in both $P_{\rm ano}$ and $1-P_{\max}$ (Figure~\ref{fig:hybrid}). The only exception is the earliest available spectrum ($+1.43$\,day after the merger), which the multi-class classifier assigns to the asteroid (Table~\ref{tab:classification}). 

\begin{figure}
    \centering
    \includegraphics[width=1\linewidth]{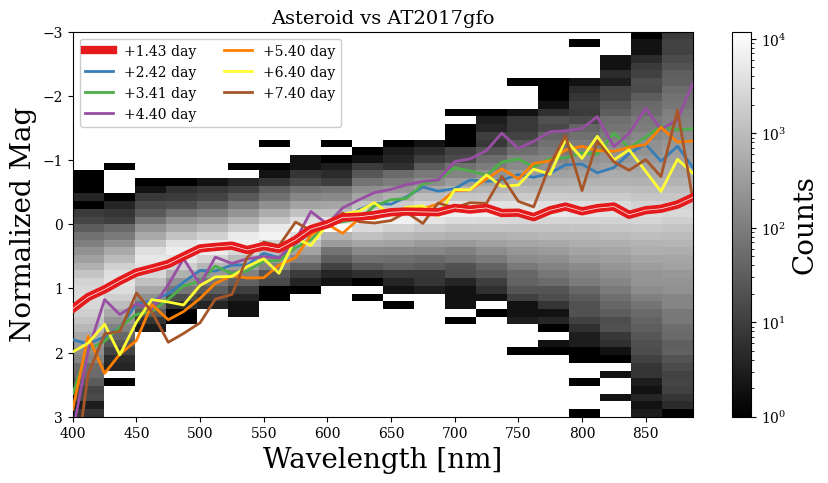}
    \caption{
    Comparison between the KN AT~2017gfo and the ensemble of all augmented asteroid samples used in training. 
    The grayscale 2D histogram shows the distribution of normalized magnitudes for asteroids across the 7DT filter set, 
    while colored lines indicate the observed SEDs of AT~2017gfo at phases +1.4 to +7.4 days post-merger. 
    }
    \label{fig:asteroid_vs_AT2017gfo}
\end{figure}

This misclassification reflects the featureless, nearly power-law continuum at early times, which resembles a solar spectrum in medium-band color space. As shown in Figure~\ref{fig:asteroid_vs_AT2017gfo}, the $+1.43$\,day spectrum overlaps the densest region of simulated asteroid SEDs, whereas later phases display slightly redder slopes that clearly separate them from the asteroid distribution. This outcome is consistent with human inspection and highlights the challenge of distinguishing ultra-blue KNe from asteroid-like continua at the earliest epochs. 

Fortunately, asteroids can be reliably excluded in practice through independent checks, such as archival matching with known solar system objects, Sky Body Tracker (\texttt{SkyBoT}; \citealt{2006ASPC..351..367B}) cross-matching, motion confirmation from sequential single exposures, and host-galaxy association. Thus, even if an early-phase KN is projected onto the asteroid locus, contextual information provides a robust discriminant.

\subsubsection{Simulated KN}\label{subsec:KNeim}
For both filter sets, the hybrid framework shows nearly identical KN recall and FPR values (Figure~\ref{fig:comp_hybrid}). This similarity arises because the broadening of spectral features due to rapid ejecta velocities and the lanthanide blanket effect \citep{2017Natur.551...80K} renders KN SEDs largely smooth and featureless. As a result, increasing the spectral resolution from 20 to 40 medium-band filters provides little additional discriminating power. 

Consequently, $\sim$90\% of simulated KNe are consistently flagged as anomalous, and $P_{ano}$ is more sensitive to the wind component than to the dynamical ejecta (Figure~\ref{fig:hybrid_kn_sim20} and ~\ref{fig:hybrid_kn_sim40}): larger wind masses produce hotter, bluer, and brighter SEDs that lie closer to the trained-class manifold, yielding lower anomaly scores, while the slowest wind case (0.05$c$) uniquely shows reduced anomaly scores, plausibly due to its delayed brightening and lower effective SNR compared to faster winds \citep{2021ApJ...918...10W}. The earliest phase (0.125~day; $\sim$3~hr) attains the highest $P_{ano}$ because its disk-driven ultra–blue, nearly featureless continuum is more extreme than any trained class and therefore falls outside the convex hull of the supervised manifold \citep{2021ApJ...918...10W}.

\begin{figure*}
    \centering
    \includegraphics[width=1.0\linewidth]{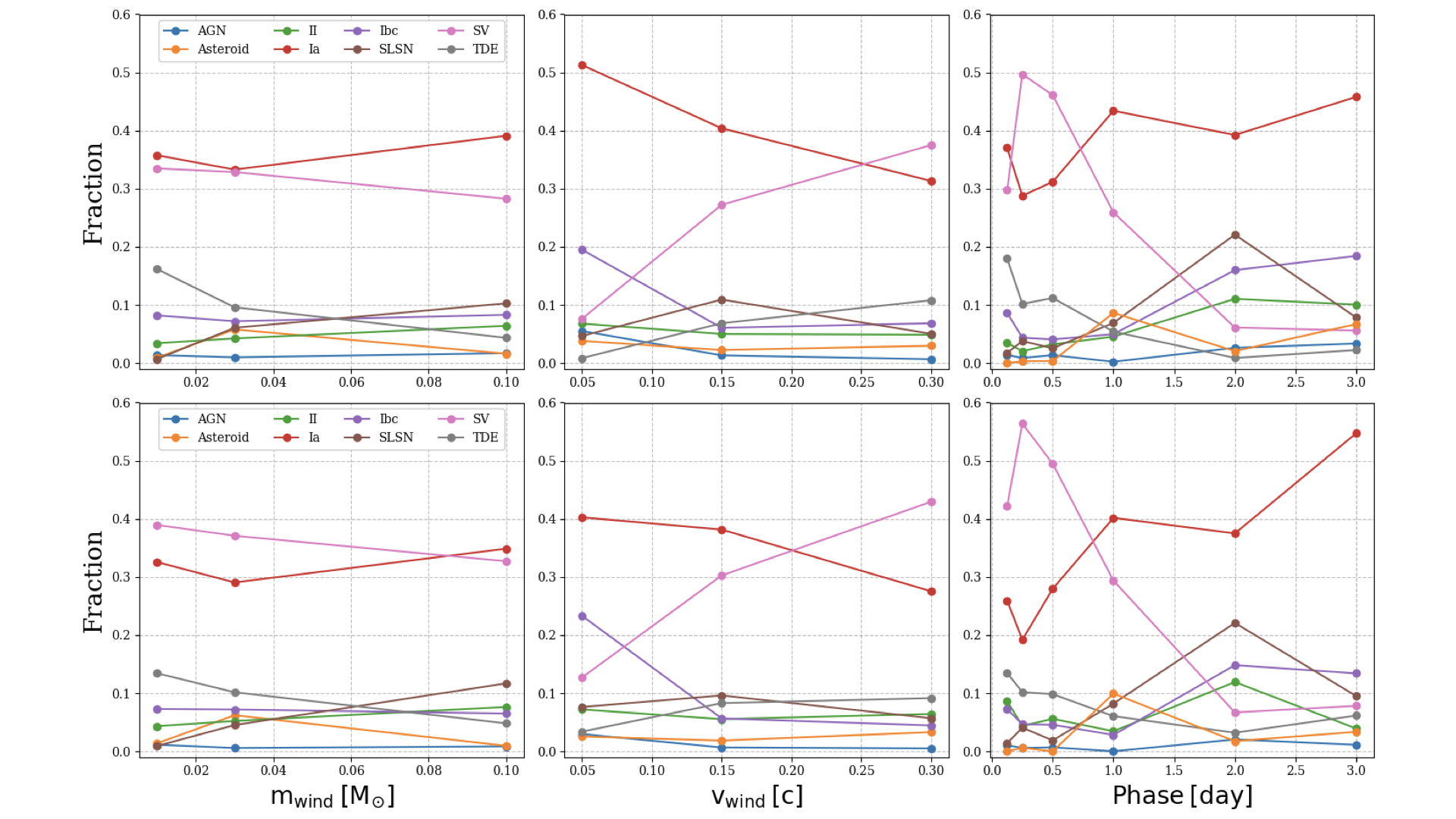}
    \caption{
    Fractions of simulated kilonovae (KN) assigned to each class by the \emph{Multi-class classifier} for varying ejecta parameters: wind mass ($m_{\mathrm{wind}}$, left), wind velocity ($v_{\mathrm{wind}}$, middle), and phase (right). 
    The top row corresponds to the 20-filter configuration and the bottom row to the 40-filter configuration. 
    Classification here is determined solely by the class with the maximum posterior probability ($P_{\max}$), irrespective of its absolute value. 
    }
    \label{fig:kn_sim_frac}
\end{figure*}

A closer inspection of the multi-class classifier results helps clarify why the anomaly score varies with KN properties. 
In Figure~\ref{fig:kn_sim_frac}, both the 20- and 40-filter configurations show that simulated KNe are most frequently misclassified as either TypeIaSNe or SV, largely independent of the wind ejecta mass.
For the wind velocity, the fraction of KNe mapped to SV increases toward higher velocities, consistent with the hotter, blue-continuum appearance of fast winds.
In contrast, the phase dependence reveals a gradual shift from SV-like classifications at early phases to Ia-like ones at later times, except for the second phase, which shows a transient increase in SV fraction.

The phase dependence exhibits a general trend in which KNe at early phases ($<$~\,0.5~\,day) are mostly mapped to SV or TDE loci, while later phases show an increasing fraction of Type Ia–like classifications. 
Around $\sim$2~\,days, however, the classifier briefly assigns a small fraction of KNe to SLSNe, Type II, and Ibc SNe, suggesting temporary color–temperature overlap during the intermediate cooling phase.
This transient shift likely reflects the evolving continuum slope as the ejecta cools, before converging toward Ia–like SEDs at $>$~\,3~\,days.


Importantly, however, all tested cases still fall within the Youden's~\,$J$ thresholds and are successfully flagged as anomalous by the hybrid framework, underscoring its robustness across the KN parameter space.


\subsubsection{Real-sky validation with 7DT}\label{subsec:7dtobs}

\begin{figure}
    \centering
    \includegraphics[width=1.0\linewidth]{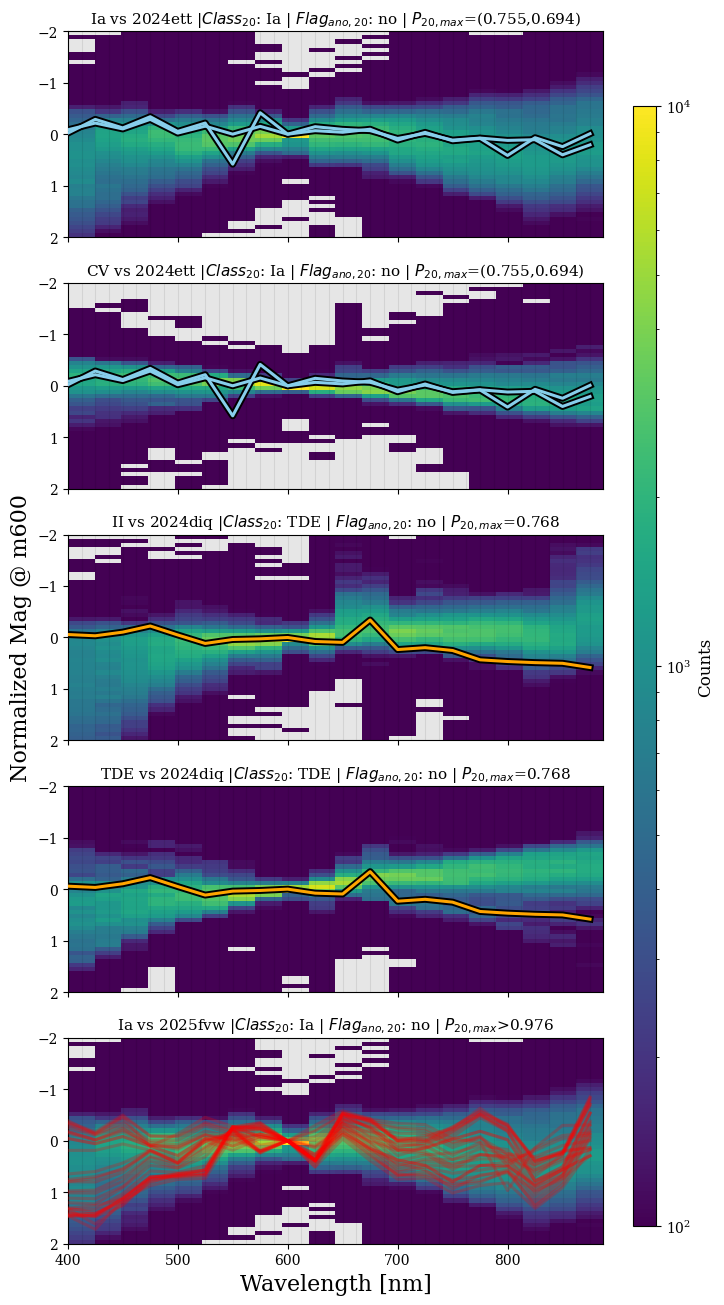}
    \caption{
    Comparison of test objects with the training+validation distributions of selected classes, shown as normalized magnitudes with respect to the $m600$ band. Each panel overlays the test SED (colored polylines; blue: AT~2024ett, orange: SN~2024diq, green: SN~2025fvw) on a 2D histogram of the corresponding class distribution. 
    Line styles indicate classification outcome (solid = correctly classified, dashed = misclassified). 
    Titles summarize the predicted class, anomaly flag, and maximum classification probability ($P_{\rm max}$). 
    }
    \label{fig:comp7dt}
\end{figure}

Applying the framework to external 7DT tests reveals both strengths and limitations. 
As shown in Figure~\,\ref{fig:comp7dt}, all phases of SN~2025fvw (Type~Ia SN) are correctly identified and lie squarely within the Ia distribution, indicating robustness for well-sampled classes. In contrast, AT~2024ett (CV) shows a CV-like shape but a pronounced $m550$ feature, placing it closer to the broad Ia locus; this is compounded by few CV examples (grouped with SV; Table~\ref{tab:7dt_val}). SN~2024diq (Type~II~SN) exhibits strong H$\alpha$ yet trends redward into a minor TDE subset, leading to confusion. These failures arise from intrinsic class overlap, training imbalance, and reduced long-wavelength sensitivity that downweights otherwise diagnostic features. Importantly, under the hybrid framework, all three objects are flagged as normal (non-anomalous), preventing erroneous rejection of common populations even when fine-grained class labels are uncertain.

\subsubsection{KN contaminant candidates from non-KN}\label{subsubsec:contaminant}

We further examine specific subclasses that have been reported as potential contaminants of KNe, particularly LBVs and Type~\,IIb SNe, which were identified as major sources of confusion in light-curve--based classifications \citep{2025PASP..137h4105B,2025MNRAS.542..541F}, to investigate which transient types could act as KN contaminants in our framework among non-KN classes.

\begin{figure*}
    \centering
    \includegraphics[width=0.8\linewidth]{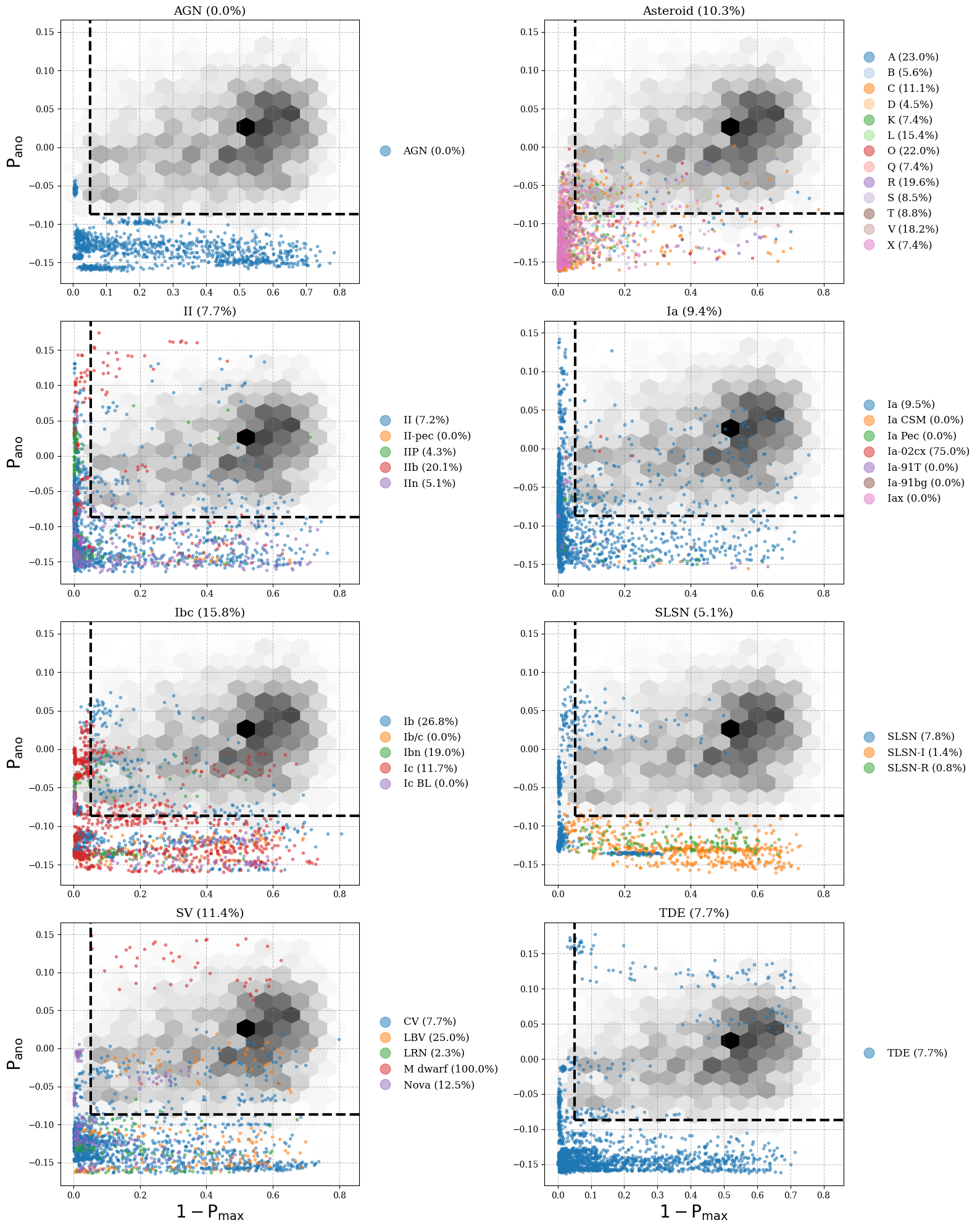}
    \caption{
    Distribution of non-KN test samples in the hybrid framework decision space for the 40 filter set configuration. 
    Each panel corresponds to one of the transient classes defined in Table~\ref{tab:type_mapping}, showing how the initially claimed subtypes from OSC and WISeREP are distributed with respect to the anomaly probability ($P_{\mathrm{ano}}$) and $1-P_{\max}$. 
    Colored points denote the subtypes redefined according to the simplified mapping used for this analysis, and the fraction in parentheses in each panel title indicates the percentage of sources classified as anomalies under the Youden's~\,$J$ threshold as described in Section~\,\ref{subsubsec:combined_decision}. 
    Legends on the right summarize the relative anomaly fractions for each subtype. 
    The gray hexagonal density background represents the distribution of simulated KNe, illustrating their characteristic region in the decision space. 
    }
    \label{fig:subtype_distribution_40}
\end{figure*}

Figure~\ref{fig:subtype_distribution_40} presents the distribution of non-KN test samples across the hybrid decision space, showing the fraction of each class that is misclassified as an anomaly under the Youden's~\,$J$ threshold.
A qualitatively consistent trend is also seen for the 20-filter configuration (Figure~\ref{fig:subtype_distribution_20}).

Among all non-KN classes, no AGN sample was misclassified as an anomaly, suggesting that the framework is robust in distinguishing KNe from AGN-like variability even when the KN occurs near galactic nuclei. 
For Type~Ibc, the contamination fraction reaches $15.8\%$, mainly driven by Type~Ib ($26.8\%$ of the Ib subsample). 
For Type~Ia, the Ia-02cx subtype shows a high anomaly fraction ($75\%$), but given the small sample size, this is not statistically significant. 
Asteroids show a wide range of anomaly ratios across more than ten subtypes. Although this work does not focus on detailed subtype-specific behavior, such false positives can be efficiently excluded in practice through contextual information--cross-matching with archival data, host-galaxy association, and multi-epoch tracking of motion.

Approximately $20\%$ of the Type~\,IIb and $25\%$ of the LBV samples were classified as anomalies--non-negligible fractions, yet somewhat reduced compared to earlier studies based solely on light-curve--based classification. 
This improvement suggests that the medium-band SED classification provides additional leverage in separating these contaminants from KNe.
By contrast, all M-dwarf flares in the test set (100\%) were flagged as anomalies. 
We note, however, that the available sample is small (15 spectra; Table~\ref{tab:type_mapping}), which limits the statistical significance of this estimate.

One noteworthy feature is that M-dwarf flares, some TDEs, and certain Type~\,II SNe exhibit distinctive distributions that overlap with simulated KNe, characterized by elevated $P_{\mathrm{ano}}>0.05$ and broadly scattered $1-P_{\max}$. 
For M-dwarf flares, this outcome is consistent with their red-dominated, featureless SEDs (see Figure~\ref{fig:synphot_example}). 
More broadly, this trend likely arises because the multi-class classifier shows ambiguity as a normal class, producing diffuse probabilities in $1-P_{\max}$, whereas the anomaly detector recognizes these sources as statistically distinct from the dominant non-KN population. 
Such behavior may reflect their intrinsically uncommon SEDs or extreme phases not well represented in the training data, resulting in both classifier ambiguity and anomaly separation. 
These interpretations suggest that the hybrid framework captures both diversity and methodological limitations, warranting further investigation.

Overall, the contamination rate remains low across the non-KN population, demonstrating that the hybrid framework effectively isolates KNe in the decision space. 
In practice, however, when a photometric candidate emerges as an anomaly, one should still consider the possibility of contamination from these rare but persistent transient classes. 
Moreover, the threshold defined by Youden's~\,$J$ statistic in this hybrid framework may not perfectly isolate KNe as clean anomalies, implying that more flexible or non-linear criteria could be required.

\subsection{Model Performance between filter sets and Class-wise Interpretation}\label{subsec:model_per}



Overall, the 40-filter configuration improved class-wise performance by $\sim$3--9.5\% relative to the 20-filter case. Nevertheless, both setups exhibit distinct behaviors across the eight transient classes in the multi-class classifier. Importantly, the marginal improvement between the 20- and 40-filter modes also extends to the anomaly classifier and the hybrid framework: both show nearly identical performance in recovering KNe as anomalies. This consistency suggests that single-epoch, medium-band SEDs already encode sufficient information to distinguish KNe from other transient populations without relying on higher spectral sampling.

As shown in Figure~\ref{fig:confusion_matrix}, asteroids consistently achieve the highest scores under both configurations. This outcome is primarily attributable to their intrinsic spectral simplicity: their SEDs are dominated by a smooth, solar-like (G2V) spectrum with only modest variations introduced by reflectance subtypes. Unlike extragalactic transients, asteroids exhibit neither temporal evolution nor redshift-induced spectral shifts. Even when accounting for scatter in brightness and template uncertainties, the simulated asteroid set remains morphologically simpler and less diverse than heterogeneous observational datasets. Consequently, the classifier identifies asteroids more readily than other classes, leading to systematically higher performance.

Type~Ia SNe follow as one of the best-performing classes. Their strong performance reflects both the abundance and diversity of training samples (Table~\ref{tab:type_mapping}) and the presence of distinctive features, such as the Si\,\textsc{ii}\,$\lambda$6355 absorption line and the absence of $H_{\alpha}$ line, within the sensitive optical range of the 7DT filters. Type~II SNe also perform well, benefiting from similarly rich training data and the strong $\rm H_{\alpha}$ emission feature. Type~Ibc SNe show comparable robustness but are confused with Type~II at the $\sim$11\% level, underscoring the difficulty of distinguishing them based solely on single-epoch medium-band data with a 20 filter set. While the 40-filter configuration improves their TPR, misclassifications with Type~II persist, suggesting that even higher spectral resolution may not fully resolve this intrinsic degeneracy.

SLSNe exhibit the lowest performance among the eight classes, a natural consequence of their rarity and the resulting paucity of training samples. Confusion with Type~II is understandable, since SLSN-II shares similar features such as blue continua and Balmer lines. Misclassifications into the stellar-variable (SV) class are also prominent: 12–14\% of true SLSNe are predicted as SV, whereas only 1–2\% of SV are mapped to SLSNe. This asymmetric confusion reflects the diverse and populous SV class, which tends to absorb SLSNe lacking strong spectral features in single-epoch SEDs. Both classes can exhibit blue continua or Balmer emission, as seen in LBVs and novae. Moreover, SLSNe are intrinsically extremely luminous-brightness alone is a strong discriminator-but our model operates on color features, so absolute luminosity is deliberately removed and cannot be exploited by the classifier. While the 40-filter configuration provides slight improvements, structural confusion remains, highlighting the need for time-domain information (slow rise and decay) and luminosity priors. With redshift or distance information from a host galaxy, however, SLSNe are easily distinguishable thanks to their extreme intrinsic luminosities.

TDEs achieve moderate, though not outstanding, performance. Notably, they exhibit $\sim$10\% confusion with AGN---consistent with previous reports of shared spectral characteristics \citep{2017ApJ...843..106B,2019MNRAS.488.4042T}. By contrast, TDEs are characterized by very broad line widths ($\sim\!10^4~\mathrm{km\,s^{-1}}$) and distinct high-ionization features such as He\,\textsc{ii}, N\,\textsc{iii}, and O\,\textsc{iii} \citep{2020MNRAS.494.2538N}, whereas AGN typically display relatively narrower emission lines with prominent narrow forbidden transitions (e.g., [O\,\textsc{iii}] and [N\,\textsc{ii}]). Nevertheless, both phenomena occur in galactic nuclei and share spectral similarities, including blue continua. Differences in continuum-color evolution and variability timescales, however, imply that single-epoch observations are often insufficient to cleanly separate the two classes; multi-epoch monitoring is required \citep{2025ApJS..281....7Y}.

Finally, although the 40-filter configuration provides incremental gains ($\sim$3\% overall), the improvements beyond the most informative bands are modest. The largest per-class benefit appears for AGN ($\sim$10\%), consistent with finer sampling (12.5\,nm steps). 
Conversely, SLSN misclassifications into Type~II/SV are amplified by limited training diversity, underscoring the structural challenge of representing rare classes. 
Possible label contamination in the archival training set cannot be ruled out and could also contribute to these limitations of the multi-class classifier.

Although the 7DT array uniquely provides simultaneous multi-band observations, the classification framework presented here is not limited to this facility. 
Any photometric system employing medium- or narrow-band filters with comparable or higher effective spectral resolution can reproduce a similar training scheme by generating synthetic photometry and color-based features. 
In such cases, the performance reported in this study can serve as a benchmark for what can be achieved with comparable filter density and wavelength coverage, regardless of the specific telescope or survey configuration.



\subsection{Optimizing Filter Configurations and Operational Strategies}\label{subsec:opt}

The results above guide observational strategies, particularly in optimizing filter configurations and contextual information under realistic resource constraints. While the 40-filter configuration achieves the highest classification accuracy, the 20-filter mode already delivers near-identical performance, with only a $\sim$3\% reduction in macro $F_{1}$ and negligible impact on anomaly detection. 

Accordingly, the 20-filter configuration represents a highly efficient operational baseline. In situations where rapid follow-up is essential---such as during GW-EM counterpart searches---prioritizing the 20-filter setup allows for alternative optimizations: securing multi-epoch coverage, extending individual exposure times, or increasing survey area per night. These strategies maximize the scientific return under limited time or hardware conditions, offering a more economical trade-off between coverage and precision. The full 40-filter configuration, in contrast, is best suited when the goal shifts from real-time triage to detailed studies of transient spectral evolution.

The filter importance analysis also offers practical guidance: when technical or maintenance constraints reduce the number of operational filters, selecting the top-ranked subset ensures minimal information loss. This approach allows survey operations to maintain high classification performance while adapting dynamically to available resources.

Finally, LSST–7DT synergy experiments show that incorporating LSST's high-SNR broad bands provides only marginal improvements ($\sim$1--2\%), with gains mainly from the $u$ and $y$ bands that extend beyond 7DT's wavelength range. This modest enhancement likely reflects both the spectral overlap and the predominance of low-$z$ samples already well constrained by 7DT colors. Nevertheless, hybrid configurations remain operationally appealing: combining SHAP-ranked top 20 medium bands of 40 filter set with a single LSST band reproduces near-ceiling performance, enabling rapid and flexible follow-up.

\subsection{Limitations}\label{subsec:limit}

We note several limitations of our findings, arising from imperfections in the training samples and bandwidth limitations of medium-band photometry.

Medium-band SEDs are good at sampling broad features of SEDs. 
However, hot, featureless, black-body dominated nature of early KNe spectra are difficult to distinguish from similarly hot, featureless spectra of early core-collapse SNe, shock-heated cooling emissions of SNe~IIb, and some of hot stellar transients such as CVs (e.g., \citealt{2025PASP..137h4105B}). 
Our spectral library contains few early SNe spectra, so our SED classifier may misclassify these imposters as KNe. 
This limitation needs to be kept in mind. 
Later epoch observations should make it possible to break this degeneracy.

Another complication arises from the dust extinction (either from host or surrounding environments) that alter the overall continuum shape. Since no attempt was made to correct for dust extinction of the training/test spectra, extinction is already reflected to the classification performance to some degree. 
Nevertheless, extensive tests with a variety of extinction condition in future can make the performance metrics of the classifier more robust.


The coarse sampling of medium-bands make it difficult to identify weak, narrow emission or absorption lines, further complicating the classification of some of the transient. 
Our training samples include such objects so our evaluation of the performance already reflects this point. 
Nevertheless, this point can be explored further in future.



Not only that, our data augmentation scheme did not fully capture the intrinsic ``diversity'' of underrepresented classes, such as SLSNe and subclasses within the SV category. 
While it alleviated small-sample and class imbalance issues by increasing the per-class sample size and maintaining uniform ratios, these synthetic variations could not reproduce the full range of physical diversity inherent to rare transients. 
Notably, injecting realistic photometric scatter helped the classifier internalize per-filter reliability consistent with the measured throughput (i.e., effectively down-weighting noisier bands), thereby improving calibration under finite SNR conditions.


These include class imbalance and data augmentation that do not fully capture the diversity in redshift and phase. 
A highly skewed redshift distribution ($>$90\% at $z<0.15$, Figure~\ref{fig:redshift}), biasing the classifier toward nearby sources.
More specifically on the redshift bias limits the model's exposure to higher-$z$ transients. 
At larger redshifts, both the overall SED color and the filter coverage of key spectral features shift systematically---not only the overall SED shape but also diagnostic emission and absorption lines move across different 7DT filters as $z$ increases. 

\begin{figure}
    \centering
    \includegraphics[width=1.0\linewidth]{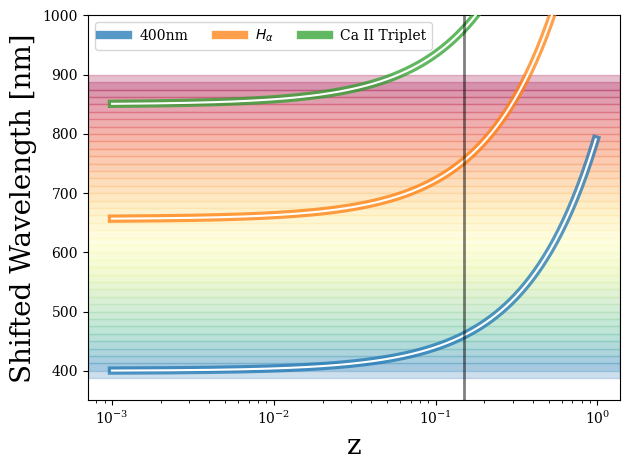}
    \caption{Redshift evolution of selected rest-frame features, shown as observed wavelengths
$\lambda_{\rm obs}=\lambda_{\rm rest}(1+z)$, overlaid on the 7DT medium-band filters
(400--887.5\,nm; 12.5\,nm spacing). Curves illustrate the drift of a continuum point at
400\,nm (blue), H$\alpha$ (656.3\,nm; orange), and the 850\,nm indicating Ca\,{\sc ii} triplet ($\sim$850\,nm; green).
The vertical line marks $z=0.15$.}
    \label{fig:z_drift}
\end{figure}

Figure~\ref{fig:z_drift} visually demonstrates this effect by showing the observed–wavelength drift of representative features: H$\alpha$ ($\lambda_{\mathrm{rest}}\!\approx\!6563$\,\AA) remains within the 7DT range up to $z\simeq0.35$, whereas the 850\,nm indicating \ion{Ca}{2} near-IR triplet ($\lambda\lambda 8498,\,8542,\,8662$\,\AA) leaves the coverage at $z\simeq0.044$, and a continuum point at 400\,nm shifts beyond 700\,nm by $z\simeq0.1$. 
Moreover, because 7DT's sensitivity declines steeply beyond $\sim$700\,nm, the classifier’s performance becomes increasingly limited for more distant, redder targets. 
This effect is largely invisible under our low-$z$ training/testing configuration; accordingly, the performance reported here should be interpreted as most representative at $z\!\lesssim\!0.15$.

As a rough detectability benchmark for common transients, a typical Type~\,Ia~\,SN with standardized peak absolute magnitude $M_r\!\approx\!-19.3$ (e.g., Pantheon assumptions; see \citealt{2021ApJ...912..150D}) would reach $m\!\sim\!20.5$, corresponding to $D_L\!\approx\!9.1\times10^{2}$\,Mpc ($z\!\sim\!0.2$). This indicates that high-$z$ contamination is plausible near the single-band detection depth, and redshift effects as well as instrumental response must be considered when interpreting feature importance and class boundaries.


When the hybrid framework was empirically tested on AT~\,2017gfo, most epochs were correctly identified as anomalous, except for the earliest phase (1.43~\,day), which was misclassified as an asteroid. 
As shown in Section~\ref{subsubsec:contaminant}, approximately 10\% of asteroids among the test samples were likewise misclassified as anomalies. 
This misclassification highlights a fundamental challenge in identifying KNe during their early phase ($<$1.5~\,day), when diagnostic spectral or color features have yet to emerge. 
Fortunately, asteroids can be reliably excluded through contextual checks, such as cross-matching with known solar-system objects using the Sky Body Tracker (\texttt{SkyBoT}; \citealt{2006ASPC..351..367B}), verifying motion across sequential exposures, or confirming host-galaxy associations. 
Thus, even if an early-phase KN lies along the asteroid locus, auxiliary information provides a robust discriminant.

The similarity between early SNe and KNe spectra also poses another challenge, but not explored extensively here due to the paucity of such early SNe spectra. 
In real observations, the chance of catching a very early SN is rare. 
Therefore, such objects will not cause too much trouble in search of KNe, although few early SNe could be identified as possible KNe during the search.

However, M-dwarf flares occasionally appeared as potential anomalies within this framework (see Section~\ref{subsubsec:contaminant}), underscoring the complexity of distinguishing rare astrophysical contaminants from true extragalactic events. 

Beyond these empirical considerations, several framework-specific limitations warrant attention. 
The thresholds and performance metrics were derived using simulated KNe detectable within the 7DT sensitivity range. 
Because the real diversity of KNe remains poorly constrained owing to the limited number of observed events, the framework may inherently reflect biases of the theoretical models, regardless of how broadly they span the physical parameter space. 
Moreover, detectable KNe in 7DT are intrinsically biased toward brighter events--those with higher wind ejecta masses or phases near peak luminosity--so faint or red KNe dominated by lanthanide-rich ejecta may still be misclassified. 

In addition, the anomaly detector (\texttt{iForest}) was trained exclusively on non-KN events, creating an implicit dependence on the distribution of these background samples. 
Since these non-KN examples themselves exhibit redshift bias, the classifier can erroneously flag distant non-KN transients as anomalies when they fall outside the manifold of the training data. 
Hence, the anomaly thresholds are better interpreted as decision-support tools that rank candidate likelihoods, rather than as rigid classification boundaries. 

Finally, an operational limitation arises from the model architecture, which assumes complete filter coverage. 
Missing observations cannot simply be treated as \texttt{NaN} inputs, because the classifier interprets non-detections as informative patterns rather than absent data. 
Consequently, incomplete filter configurations would require retraining to preserve feature interpretation, posing a practical challenge for real-time applications.

Additionally, the heterogeneous archival spectra with inconsistent resolutions and calibrations; 
and compounded uncertainties when generating synthetic photometry (original spectral errors plus filter-throughput weighting). 
These considerations argue for more homogeneous and diverse training data spanning a broader redshift and phase range, as well as redshift-aware training.

The rapid increase in spectroscopic samples expected in the LSST era, combined with theoretical models, will enable the construction of more balanced training sets spanning redshift, phase, and transient type more uniformly. 
However, reliance on theoretical models carries the risk of introducing biases if their assumptions deviate from reality, requiring caution when integrating such synthetic data.

Finally, this study is extendable to similar systems having multiple narrow/medium-bands in optical wavelength. 
\texttt{7DT-Simulator} is a flexible tool to compose the user-input sequential bands and to generate synthetic photometries varying the bandwidth.
Therefore, it is useful to design a telescope system to predict and optimize the classification power as well as training classifier for its own one.

\subsection{Future Works}\label{subsec:future}
Looking ahead, several directions for improvement are apparent. 

First, although this work demonstrates that single-epoch SEDs can already deliver robust performance and efficiency, further gains can be obtained by adding temporal information. 
For example, experiments with two-epoch configurations can allow us to evaluate what time gaps (e.g., hours to days) provide optimal leverage for classification. 
Such an approach naturally extends to Rubin synergy tests, where realistic alert cadences can be incorporated to assess time-critical follow-up strategies that account for transient evolution.

Second, our analysis has so far focused on color-only features, without explicitly incorporating contextual information such as host-galaxy properties.
Recent studies (e.g., \citealt{2024arXiv241017322J}) demonstrate that host-galaxy SEDs, redshifts, and local environments can substantially aid transient classification. 
Incorporating such contextual priors into the 7DT framework could further reduce degeneracies, particularly between transients themselves or transients with distinct host-galaxy preferences.

Finally, mitigating redshift-dependent feature drift and declining throughput at $\gtrsim700$\,nm will require both a more balanced redshift distribution in the training data and instrumental extensions toward the NIR. 
Augmenting 7DT with more sensitive NIR-capable cameras, or synergizing with complementary facilities, would enhance the classification of redshifted or late-phase transients such as high-$z$ SNe and red KNe.

\section{Conclusion}\label{sec:conclusion}

We developed and validated a 7DT-specific, single-epoch classification framework designed to emulate realistic survey operations, coupling a supervised multi-class classifier with an unsupervised anomaly detector to identify KNe amid common transients using medium-band SEDs without using KNe in training. 
This design targets time-critical follow-up by avoiding multi-epoch light-curve requirements, while remaining interpretable and operationally simple.

Across eight transient classes, the classifier attains macro $F_{1}\!\sim\!0.80$ (20-filter set)–0.82 (40-filter set). 
In combination with the anomaly detector, optically detectable KN recall exceeds $90\%$ for simulated/observed cases (including AT~2017gfo), while the non-KN FPR is reduced to $\sim$10\%. 
Thus, the hybrid approach both preserves sensitivity to KNe and suppresses contamination, making it directly useful for GW follow-up triage.
However, we note that the apparent separability the sampled phase and redshift distributions of the available training and evaluation data ($z < 0.15$ for SNe, less early-phase SNe, etc).

Filter-importance analyses (via SHAP) show that near-baseline accuracy is retained with only $\sim$40–50\% of the most informative bands, especially those sampling H$\alpha$ and Si\,\textsc{ii}. 
This finding motivates a flexible observing strategy where a minimal yet optimized subset of filters can be used during rapid-response operations. 
Consistent with this, adding a single LSST band to a top-ranked subset of 7DT filters reproduces the full 40-filter performance to within $\sim$1–2\%, demonstrating strong synergy between medium- and broad-band observations under limited resources.

The framework is computationally light---after training, thousands ($\gtrsim10^{3}$) of sources can be evaluated in seconds---facilitating integration with high-volume alert streams. 
Thresholds derived from Youden's~\,$J$ are intended as prioritization guides rather than hard cutoffs, supporting ranked follow-up decisions in real time.

These gains come with recognizable limits shaped by realistic operations: 
(i) training and simulation sets remain biased in phase and redshift, especially lacking faint or high-$z$ examples; 
(ii) throughput-weighted systematics arise from synthetic photometry; and 
(iii) additional uncertainty from heterogeneous archival spectra. 

Our low-resolution spectro-photometric classification presents both advantages and disadvantages in comparison to other conventional classifications; light curve-based classification and spectroscopy.
Complementary contaminant tests indicate practical resilience: AGN-like variability is cleanly rejected, and traditional photometric confusers such as LBVs and Type\,~IIb~\,SNe are substantially mitigated by the medium-band SED information. 
By contrast, M-dwarf flares can still be flagged as anomalies--a residual weakness of the hybrid framework--although the available test set for this class is limited in diversity. 

In practice, conservative thresholds, compact SHAP-guided filter menus, and a single broad-band anchor (when available) hedge against these risks until more diverse observations accumulate, while coupling the ranked anomaly list with lightweight contextual features can further reduce residual contamination in real-time operations.

In summary, 7DT's medium-band system, coupled with the hybrid classifier–anomaly framework, delivers fast, interpretable, and scalable transient vetting from single-epoch data. 
This capability is well matched to KN searches in the LSST era and to the growing cadence and volume of next-generation GW-EM follow-up.
\begin{acknowledgments}

We acknowledge the support from the National Research Foundation of Korea (NRF) grant funded by the Korean government (MSIT), NRF-2021M3F7A1084525, the GW Universe project.

G.S.H.P. acknowledges support from the Pan-STARRS project, which is a project of the Institute for Astronomy of the University of Hawai'i, and is supported by the NASA SSO Near Earth Observation Program under grants 80NSSC18K0971, NNX14AM74G, NNX12AR65G, NNX13AQ47G, NNX08AR22G, 80NSSC21K1572, and by the State of Hawai'i.

S.W.C. acknowledges support from the National Research Foundation of Korea (NRF) grants, No. 2021M3F7A1084525 funded by the Ministry of Science and ICT (MSIT). This research was also supported by Basic Science Research Program through the NRF funded by the Ministry of Education (RS-2023-00245013). 

J.H.K also acknowledges the support from the Institute of Information \& Communications Technology Planning \& Evaluation (IITP) grant, No. RS-2021-II212068 funded by the Korean government (MSIT).

HC acknowledges support from the National Research Foundation of Korea (NRF) grant (RS-2025-00573214) funded by the Korea government (MSIT).

Data transfer from the host site was supported in part by the Korea Research Environment Open NETwork (KREONET) advanced research program funded by KISTI.

This work was supported by resources provided by University of Hawai'i Information Technology Services – Research Cyberinfrastructure (ITS-RCI), funded in part by the National Science Foundation CC* awards \#2201428 and \#2232862, including use of the Koa HPC cluster.

We also thank Eugene Magnier (UH IfA) for general discussion about the result, Dongin Lee (POSTECH) for valuable discussions regarding machine learning experiments, and Sophia Kim (SNU ARC) for insightful advice on interpreting SN spectra.

\end{acknowledgments}


\begin{contribution}
\textbf{Conceptualization:} G.S.H.P. (lead)\\
\textbf{Methodology:} G.S.H.P. (lead)\\
\textbf{Software:} G.S.H.P. (lead)\\
\textbf{Validation:} G.S.H.P.\\
\textbf{Formal analysis:} G.S.H.P.\\
\textbf{Investigation:} G.S.H.P.\\
\textbf{Data curation:} G.S.H.P. (lead), H.C. (support)\\
\textbf{Resources:} G.S.H.P.; (7DT telescope system) M.I., J.K.\\
\textbf{Project administration:} G.S.H.P.; (7DT telescope system) M.I., J.K.\\
\textbf{Supervision:} M.I.\\
\textbf{Funding acquisition:} M.I., G.S.H.P.\\
\textbf{Visualization:} G.S.H.P.\\
\textbf{Writing – original draft:} G.S.H.P.\\
\textbf{Writing – review \& editing:} G.S.H.P., S.-W.C., M.I., J.K., H.C.\\
\end{contribution}



%

\software{
    Astropy \citep{2013A&A...558A..33A,2018AJ....156..123A,2022ApJ...935..167A},
    NumPy \citep{2020Natur.585..357H},
    SciPy \citep{2020NatMe..17..261V},
    Source Extractor \citep{1996A&AS..117..393B},
    SWarp \citep{2010ascl.soft10068B},
    SCAMP \citep{2006ASPC..351..112B},
    XGBoost \citep{2016arXiv160302754C},
    LightGBM \citep{10.5555/3294996.3295074},
    CatBoost \citep{2018arXiv181011363V},
    Isolation Forest \citep{Liu2012}.
}


\appendix

\section{Synthetic Asteroid Spectra}\label{app:asteroid}

To evaluate the classifier's ability to reject asteroids, we constructed synthetic asteroid spectra following the Bus--DeMeo taxonomy \citep{2009Icar..202..160D}. 
We adopted the publicly available Bus--DeMeo mean reflectance spectra, which provide both the class-averaged reflectance and the associated 1$\sigma$ dispersion across the 0.45--2.45~$\mu$m range for 25 asteroid classes: \{A, B, C, Cb, Cg, Cgh, Ch, D, K, L, O, Q, R, S, Sa, Sq, Sr, Sv, T, V, X, Xc, Xe, Xk, Xn\}. 
Each class is characterized by the average spectral behavior of known members within that group. 

To cover the wavelength range relevant for our system response modeling (1000--25000~\AA), we extrapolated the reflectance curves using a hybrid Gaussian Process (GP) regression framework. 
For the blue end, we fitted a first-order polynomial to the bluest 3--5 points of each spectrum, generating linear anchor points down to 1000~\AA. 
These anchors were combined with the original data as priors for GP training. 
The resulting GP models provided smooth, probabilistic extrapolations with uncertainties, from which we sampled random reflectance realizations to capture intrinsic diversity. 

To simulate realistic observations, each reflectance realization was multiplied by a high-resolution solar spectrum from the \textit{HST} CALSPEC archive\footnote{\url{https://www.stsci.edu/hst/instrumentation/reference-data-for-calibration-and-tools/astronomical-catalogs/solar-system-objects-spectra}}, specifically the empirical solar spectrum constructed from \citet{2003SoPh..214....1T} for 0.2–2.4~$\mu$m combined with a local thermal equilibrium (LTE) model \citep{1974SoPh...39...19H} at longer wavelengths. 
The spectrum was normalized to a heliocentric distance of 1~\,AU and interpolated onto a regular wavelength grid matching our required resolution. 

The synthetic asteroid spectra were then normalized to the solar flux at 5500~\AA\ and scaled to a fixed reference level ($10^{-15}$ erg~s$^{-1}$~cm$^{-2}$~\AA$^{-1}$). 
We further introduced apparent magnitude variations by sampling from an exponential magnitude distribution with a tunable slope parameter $\alpha$, yielding a realistic brightness distribution peaking near the survey's detection limit. 
Flux scaling was applied via flux–magnitude conversion at 5500~\AA, producing spectra spanning a representative range of apparent magnitudes. 

\begin{figure}
    \centering
    \includegraphics[width=0.5\linewidth]{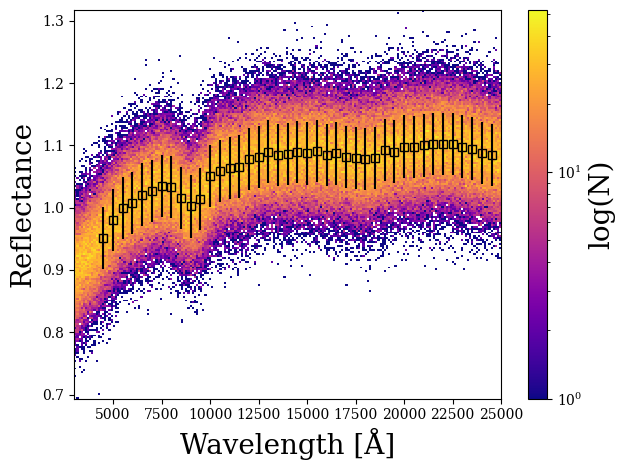}
    \caption{Example of GP-based extrapolation and synthetic spectra generation of S-type asteroid from \citealt{2009Icar..202..160D}.}
    \label{fig:app_stype_asteroid}
\end{figure}

For each of the 25 asteroid types, we generated $N=1000$ synthetic spectra, resulting in a total of 25,000 samples. 
Figure~\ref{fig:app_stype_asteroid} shows an example of the GP extrapolation, reflectance ensemble generation, and resulting flux and AB magnitude spectra for an S-type asteroid. 

\section{Model Hyperparameters for Multi-class classifier}\label{app:hyperparameter}
We summarize the hyperparameters for the tests in Tables~\ref{tab:result20} and \ref{tab:result40}. 
The optimized hyperparameters obtained from the Bayesian search using \texttt{Optuna} for each model, \texttt{LightGBM}, \texttt{CatBoost}, and \texttt{XGBoost} for the classifiers trained on the 20- and 40-filter sets.  
We fixed the random seed (\texttt{random\_seed}$=$42) to ensure reproducibility and repeat 100 times with early stopping with 30 rounds based on the validation log-loss metric to maximize the macro $F_{1}$ score.

Below configurations correspond to the best-performing \texttt{CatBoost} models using full filter sets after 100 optimization trials:

\begin{itemize}
    \item \textbf{20-filter set:} 
    \texttt{depth} = 5, 
    \texttt{iterations} = 1640, 
    \texttt{learning\_rate} = 0.0509, 
    \texttt{l2\_leaf\_reg} = 3, 
    \texttt{border\_count} = 154.

    \item \textbf{40-filter set:} 
    \texttt{depth} = 5, 
    \texttt{iterations} = 1471, 
    \texttt{learning\_rate} = 0.0519, 
    \texttt{l2\_leaf\_reg} = 4, 
    \texttt{border\_count} = 129.
\end{itemize}

Below configurations correspond to the best-performing \texttt{LightGBM} models using full filter sets after 100 optimization trials:

\begin{itemize}
    \item \textbf{20-filter set:} 
    \texttt{num\_leaves} = 25, 
    \texttt{n\_estimators} = 391, 
    \texttt{min\_child\_weight} = 17, 
    \texttt{learning\_rate} = 0.0497, 
    \texttt{colsample\_bytree} = 0.7467.

    \item \textbf{40-filter set:} 
    \texttt{num\_leaves} = 18, 
    \texttt{n\_estimators} = 424, 
    \texttt{min\_child\_weight} = 15, 
    \texttt{learning\_rate} = 0.0882, 
    \texttt{colsample\_bytree} = 0.8007.
\end{itemize}

The hyperparameters for various tests using \texttt{XGBoost} are summrized in Table~\,\ref{tab:hyperparameter}.
\begin{longrotatetable}
    \begin{deluxetable*}{cccccccccc}
        \tablecaption{Model hyperprameters for multi-class classifier}
        \label{tab:hyperparameter}
        \tablehead{
        \colhead{Test Name} & \colhead{Filter set} & \colhead{\texttt{colsample\_bytree}} & \colhead{\texttt{gamma}} & \colhead{\texttt{learning\_rate}} &
        \colhead{\texttt{max\_depth}} & \colhead{\texttt{min\_child\_weight}} & \colhead{\texttt{n\_estimators}} & \colhead{\texttt{reg\_lambda}} & \colhead{\texttt{subsample}}
        }
        \startdata
        Base                 & 20 & 0.6480 & 0.2464 & 0.1533 & 8 & 7 & 1832 & 68.1640 & 0.6023 \\
        Subset Top10         & 20 & 0.6899 & 0.0197 & 0.0657 & 12 & 3 & 936 & 0.0635 & 0.6621 \\
        Subset Bottom10      & 20 & 0.7214 & 0.2249 & 0.0178 & 12 & 8 & 429 & 39.2580 & 0.6517 \\
        7DT+Rubin \textit{u} & 20 & 0.6794 & 0.0811 & 0.1853 & 6 & 8 & 1231 & 52.1295 & 0.7638 \\
        7DT+Rubin \textit{g} & 20 & 0.8319 & 0.2926 & 0.0434 & 6 & 1 & 1145 & 97.9660 & 0.7430 \\
        7DT+Rubin \textit{r} & 20 & 0.8663 & 0.1408 & 0.0231 & 8 & 8 & 980 & 93.6207 & 0.7898 \\
        7DT+Rubin \textit{i} & 20 & 0.7541 & 0.3415 & 0.0531 & 7 & 2 & 1586 & 93.3374 & 0.7539 \\
        7DT+Rubin \textit{z} & 20 & 0.8909 & 0.4640 & 0.0869 & 8 & 9 & 1089 & 97.1540 & 0.9743 \\
        7DT+Rubin \textit{y} & 20 & 0.6606 & 0.1400 & 0.0707 & 8 & 9 & 1202 & 98.4181 & 0.6123 \\
        Subset 10\%          & 20 & 0.8934 & 0.2116 & 0.0186 & 12 & 5 & 1278 & 30.5799 & 0.7041 \\
        Subset 25\%          & 20 & 0.8844 & 0.0005 & 0.0240 & 11 & 9 & 1637 & 0.3926 & 0.6719 \\
        Subset 50\%          & 20 & 0.8114 & 0.3578 & 0.0514 & 10 & 3 & 1170 & 46.0932 & 0.6976 \\
        Subset 75\%          & 20 & 0.6273 & 0.4621 & 0.1045 & 8 & 10 & 1628 & 78.6712 & 0.6847 \\
        Subset 90\%          & 20 & 0.6733 & 0.2780 & 0.0348 & 8 & 6 & 1468 & 93.3742 & 0.6572 \\
        Base                 & 40 & 0.6847 & 0.0569 & 0.1604 & 5 & 2 & 500 & 46.9499 & 0.6966 \\
        7DT+Rubin \textit{u} & 40 & 0.6748 & 0.2573 & 0.0524 & 3 & 5 & 898 & 2.3158 & 0.6310 \\
        7DT+Rubin \textit{g} & 40 & 0.7713 & 0.3020 & 0.1507 & 5 & 4 & 1857 & 99.4494 & 0.9277 \\
        7DT+Rubin \textit{r} & 40 & 0.7233 & 0.2380 & 0.2238 & 4 & 10 & 1651 & 4.9624 & 0.8989 \\
        7DT+Rubin \textit{i} & 40 & 0.6777 & 0.2303 & 0.0750 & 4 & 9 & 1837 & 24.3095 & 0.8220 \\
        7DT+Rubin \textit{z} & 40 & 0.7964 & 0.1585 & 0.2989 & 5 & 2 & 1820 & 41.4343 & 0.6934 \\
        7DT+Rubin \textit{y} & 40 & 0.9453 & 0.2557 & 0.0771 & 6 & 10 & 1987 & 3.5406 & 0.7029 \\
        7DT+Rubin \textit{u} $^\dagger$ & 40 & 0.7104 & 0.0982 & 0.0473 & 5 & 10 & 1296 & 31.1720 & 0.7504 \\
        7DT+Rubin \textit{g} $^\dagger$ & 40 & 0.6868 & 0.4574 & 0.0647 & 11 & 1 & 843 & 65.3050 & 0.6383 \\
        7DT+Rubin \textit{r} $^\dagger$ & 40 & 0.7342 & 0.1264 & 0.0207 & 12 & 1 & 1723 & 76.1291 & 0.6526 \\
        7DT+Rubin \textit{i} $^\dagger$ & 40 & 0.8539 & 0.0012 & 0.1401 & 6 & 9 & 498 & 47.0467 & 0.7290 \\
        7DT+Rubin \textit{z} $^\dagger$ & 40 & 0.6950 & 0.0521 & 0.0304 & 8 & 5 & 1354 & 22.1057 & 0.6091 \\
        7DT+Rubin \textit{y} $^\dagger$ & 40 & 0.6387 & 0.2844 & 0.0629 & 5 & 3 & 1854 & 0.0111 & 0.8342 \\
        Subset Top20         & 40 & 0.7342 & 0.3328 & 0.0280 & 7 & 5 & 1312 & 63.2550 & 0.6614 \\
        Subset Bottom20      & 40 & 0.6027 & 0.3108 & 0.0174 & 12 & 10 & 1569 & 0.0070 & 0.6797 \\
        Subset 10\%          & 40 & 0.7953 & 0.3664 & 0.0400 & 8 & 8 & 1855 & 0.0096 & 0.7649 \\
        Subset 20\%          & 40 & 0.9205 & 0.2190 & 0.1004 & 5 & 5 & 1165 & 0.3251 & 0.8422 \\
        Subset 25\%          & 40 & 0.6699 & 0.0165 & 0.1490 & 6 & 3 & 1699 & 18.5603 & 0.8418 \\
        Subset 30\%          & 40 & 0.7806 & 0.0365 & 0.0318 & 8 & 1 & 1794 & 91.6069 & 0.8034 \\
        Subset 40\%          & 40 & 0.6130 & 0.2357 & 0.0373 & 8 & 10 & 1541 & 55.5197 & 0.6353 \\
        Subset 50\%          & 40 & 0.7330 & 0.0129 & 0.0482 & 5 & 2 & 1577 & 0.0792 & 0.6805 \\
        Subset 75\%          & 40 & 0.9791 & 0.2733 & 0.1120 & 4 & 9 & 398 & 0.9625 & 0.8556 \\
        Subset 90\%          & 40 & 0.8114 & 0.4120 & 0.0904 & 4 & 6 & 1051 & 74.9603 & 0.6817 \\
        \enddata
    \end{deluxetable*}
\end{longrotatetable}


\section{7DT Medium-band Photometries}\label{app:7dtobs}
This section summarizes the photometries observed by 7DT with the 20-filter configuration. 
These data include the calibrated single-epoch SEDs of representative transients (SN~2025fvw, SN~2024diq, and AT~2024ett) obtained in the spectroscopic mode described in Section~\ref{sec:data}. 
Each photometric point corresponds to a 3~$\times$~100~s exposure per filter, processed with standard bias, dark, flat, astrometric, and photometric calibrations using the \texttt{gpPy-GPU} pipeline. 
The resulting AB magnitudes are listed in Table~\ref{tab:sn2025fvw_wide} and shown in Figure~\ref{fig:7dtobs_all}. 
These 7DT photometries serve as real-world test cases for evaluating the trained multi-class classifier and anomaly detection framework.

\begin{sidewaystable*}
\centering
\caption{7DT Medium-band Photometry (20 filters).}
\label{tab:sn2025fvw_wide}
\scriptsize
\setlength{\tabcolsep}{3pt}
\renewcommand{\arraystretch}{0.95}

\resizebox{\textheight}{!}{%
\begin{tabular}{lc|cccccccccccccccccccc}
\toprule
 Name & $\Delta t$ & $m400$ & $m425$ & $m450$ & $m475$ & $m500$ & $m525$ & $m550$ & $m575$ & $m600$ & $m625$ & $m650$ & $m675$ & $m700$ & $m725$ & $m750$ & $m775$ & $m800$ & $m825$ & $m850$ & $m875$ \\
      & (day)      & (mag)  & (mag)  & (mag)  & (mag)  & (mag)  & (mag)  & (mag)  & (mag)  & (mag)  & (mag)  & (mag)  & (mag)  & (mag)  & (mag)  & (mag)  & (mag)  & (mag)  & (mag)  & (mag)  & (mag) \\
\midrule
 SN 2025fvw & 5.5 & 14.6 & 14.9 & 14.6 & 15.1 & 14.9 & 14.6 & 14.7 & 14.7 & 15.0 & 14.7 & 14.4 & 14.6 & 14.7 & 14.7 & 15.0 & 15.0 & 16.0 & 15.1 & 14.7 & 14.7 \\
            & 7.4 & 14.2 & 14.4 & 14.1 & 14.5 & 14.4 & 14.2 & 14.3 & 14.2 & 14.5 & 14.3 & 14.0 & 14.2 & 14.3 & 14.3 & 14.6 & 14.4 & 14.9 & 14.7 & 14.5 & 14.4 \\
            & 8.4 & 14.0 & 14.2 & 14.0 & 14.3 & 14.3 & 14.0 & 14.2 & 14.0 & 14.3 & 14.2 & 13.9 & 14.0 & 14.2 & 14.2 & 14.4 & 14.3 & 14.6 & 14.6 & 14.4 & 14.3 \\
            & 9.4 & 13.9 & 14.0 & 13.8 & 14.2 & 14.2 & 13.9 & 14.1 & 13.9 & 14.1 & 14.1 & 13.8 & 14.0 & 14.1 & 14.1 & 14.4 & 14.2 & 14.4 & 14.5 & 14.3 & 14.3 \\
            & 10.4 & 13.8 & 14.0 & 13.8 & 14.0 & 14.1 & 13.8 & 14.0 & 13.8 & 14.0 & 14.1 & 13.7 & 13.9 & 14.0 & 14.1 & 14.4 & 14.2 & 14.3 & 14.5 & 14.2 & 14.1 \\
            & 11.4 & 13.7 & 13.9 & 13.7 & 14.0 & 14.0 & 13.8 & 13.9 & 13.8 & 13.9 & 14.0 & 13.6 & 13.9 & 14.0 & 14.1 & 14.4 & 14.1 & 14.2 & 14.5 & 14.2 & 14.1 \\
            & 12.4 & 13.7 & 13.8 & 13.7 & 13.9 & 14.0 & 13.7 & 13.9 & 13.7 & 13.8 & 14.0 & 13.6 & 13.9 & 14.1 & 14.1 & 14.4 & 14.2 & 14.1 & 14.5 & 14.2 & 14.0 \\
            & 13.4 & 13.7 & 13.8 & 13.6 & 13.8 & 14.0 & 13.7 & 13.8 & 13.7 & 13.7 & 14.0 & 13.6 & 13.9 & 14.1 & 14.2 & 14.5 & 14.2 & 14.2 & 14.6 & 14.3 & 14.0 \\
            & 14.4 & 13.7 & 13.8 & 13.6 & 13.8 & 14.0 & 13.7 & 13.8 & 13.6 & 13.7 & 14.0 & 13.5 & 13.9 & 14.1 & 14.2 & 14.6 & 14.2 & 14.2 & 14.6 & 14.3 & 14.0 \\
            & 16.4 & 13.8 & 13.8 & 13.7 & 13.8 & 14.0 & 13.7 & 13.8 & 13.6 & 13.6 & 14.0 & 13.5 & 13.8 & 14.1 & 14.2 & 14.6 & 14.3 & 14.3 & 14.6 & 14.4 & 13.9 \\
            & 23.3 & 14.6 & 14.6 & 14.4 & 14.2 & 14.4 & 14.0 & 14.0 & 14.2 & 14.0 & 14.5 & 14.0 & 14.4 & 14.7 & 14.7 & 15.0 & 14.9 & 14.8 & 15.3 & 14.9 & 14.2 \\
            & 24.4 & 14.9 & 14.8 & 14.6 & 14.3 & 14.5 & 14.1 & 14.0 & 14.3 & 14.1 & 14.5 & 14.0 & 14.4 & 14.7 & 14.8 & 15.0 & 14.9 & 14.6 & 15.4 & 15.0 & 14.1 \\
            & 25.4 & 14.9 & 14.9 & 14.7 & 14.4 & 14.6 & 14.2 & 14.0 & 14.4 & 14.2 & 14.6 & 14.0 & 14.5 & 14.8 & 14.8 & 15.0 & 15.0 & 14.8 & 15.4 & 14.9 & 14.1 \\
            & 26.4 & 15.1 & 15.1 & 14.8 & 14.5 & 14.7 & 14.3 & 14.0 & 14.5 & 14.2 & 14.6 & 14.0 & 14.5 & 14.9 & 14.8 & 15.0 & 14.9 & 14.8 & 15.4 & 14.9 & 14.1 \\
            & 27.4 & 15.3 & 15.2 & 15.0 & 14.6 & 14.7 & 14.4 & 14.1 & 14.5 & 14.3 & 14.7 & 14.0 & 14.5 & 14.9 & 14.8 & 14.9 & 14.9 & 14.7 & 15.3 & 15.0 & 14.1 \\
            & 28.3 & 15.4 & 15.4 & 15.1 & 14.7 & 14.8 & 14.5 & 14.1 & 14.6 & 14.3 & 14.7 & 14.0 & 14.4 & 14.9 & 14.8 & 14.8 & 14.8 & 14.7 & 15.3 & 15.0 & 14.1 \\
            & 29.3 & 15.6 & 15.6 & 15.3 & 14.8 & 14.9 & 14.7 & 14.2 & 14.6 & 14.4 & 14.8 & 14.1 & 14.5 & 14.9 & 14.8 & 14.8 & 14.7 & 14.7 & 15.3 & 15.0 & 14.1 \\
            & 30.3 & 15.7 & 15.7 & 15.4 & 15.0 & 15.0 & 14.8 & 14.2 & 14.6 & 14.5 & 14.8 & 14.1 & 14.5 & 14.9 & 14.8 & 14.7 & 14.7 & 14.8 & 15.4 & 15.0 & 14.1 \\
            & 31.5 & 15.8 & 15.8 & 15.5 & 15.1 & 15.1 & 14.9 & 14.3 & 14.6 & 14.5 & 14.9 & 14.1 & 14.4 & 14.9 & 14.8 & 14.7 & 14.6 & 14.7 & 15.3 & 15.1 & 14.1 \\
            & 33.5 & 16.1 & 16.1 & 15.7 & 15.3 & 15.3 & 15.2 & 14.4 & 14.6 & 14.6 & 15.0 & 14.1 & 14.4 & 14.9 & 14.8 & 14.6 & 14.5 & 14.6 & 15.4 & 15.2 & 14.1 \\
            & 34.3 & 16.1 & 16.2 & 15.9 & 15.4 & 15.3 & 15.3 & 14.4 & 14.6 & 14.7 & 15.1 & 14.2 & 14.4 & 14.8 & 14.8 & 14.6 & 14.5 & 14.6 & 15.4 & 15.2 & 14.1 \\
            & 36.3 & 16.3 & 16.3 & 16.0 & 15.6 & 15.5 & 15.5 & 14.6 & 14.7 & 14.9 & 15.2 & 14.3 & 14.5 & 14.9 & 14.9 & 14.7 & 14.4 & 14.7 & 15.3 & 15.1 & 14.2 \\
            & 37.3 & 16.4 & 16.4 & 16.1 & 15.7 & 15.6 & 15.6 & 14.7 & 14.7 & 15.0 & 15.3 & 14.4 & 14.6 & 14.9 & 14.9 & 14.7 & 14.4 & 14.6 & 15.4 & 15.2 & 14.2 \\
            & 40.3 & 16.5 & 16.6 & 16.4 & 16.0 & 15.9 & 15.8 & 15.0 & 15.0 & 15.2 & 15.6 & 14.7 & 14.9 & 15.2 & 15.2 & 15.0 & 14.7 & 14.9 & 15.6 & 15.5 & 14.5 \\
            & 41.3 & 16.8 & 16.6 & 16.5 & 16.1 & 16.0 & 16.0 & 15.0 & 15.0 & 15.3 & 15.7 & 14.8 & 15.0 & 15.3 & 15.3 & 15.1 & 14.8 & 15.0 & 15.7 & 15.6 & 14.5 \\
            & 42.3 & 16.7 & 16.8 & 16.5 & 16.1 & 16.1 & 16.0 & 15.1 & 15.1 & 15.4 & 15.8 & 14.9 & 15.1 & 15.4 & 15.4 & 15.2 & 14.8 & 15.2 & 16.0 & 15.7 & 14.7 \\
            & 44.3 & 16.8 & 17.0 & 16.5 & 16.3 & 16.1 & 16.1 & 15.3 & 15.3 & 15.4 & 15.9 & 15.1 & 15.3 & 15.7 & 15.5 & 15.3 & 14.9 & 15.2 & 16.4 & 15.9 & 14.9 \\
\midrule
 SN 2024diq & 1.97 & 15.8 & 15.8 & 15.8 & 15.7 & 15.8 & 16.0 & 15.9 & 15.9 & 15.9 & 16.0 & 16.0 & 15.5 & 16.1 & 16.1 & 16.1 & 16.3 & 16.3 & 16.4 & 16.4 & 16.5 \\
\midrule
 AT 2024ett & 0.254 & 16.1 & 16.0 & 16.1 & 15.9 & 16.2 & 16.1 & 16.2 & 16.1 & 16.2 & 16.2 & 16.2 & 16.1 & 16.3 & 16.2 & 16.4 & 16.3 & 16.6 & 16.3 & 16.5 & 16.2 \\
 AT 2024ett & 0.306 & 17.0 & 16.8 & 16.9 & 16.7 & 16.9 & 16.8 & 17.6 & 16.6 & 17.0 & 16.9 & 16.9 & 17.0 & 17.1 & 17.0 & 17.1 & 17.0 & 17.0 & 17.3 & 17.3 & 17.1 \\
\bottomrule
\end{tabular}
} 
\end{sidewaystable*}

\section{Result of Hybrid framework}\label{app:hybrid20}

\begin{figure*}
    \centering
    \includegraphics[width=1\linewidth]{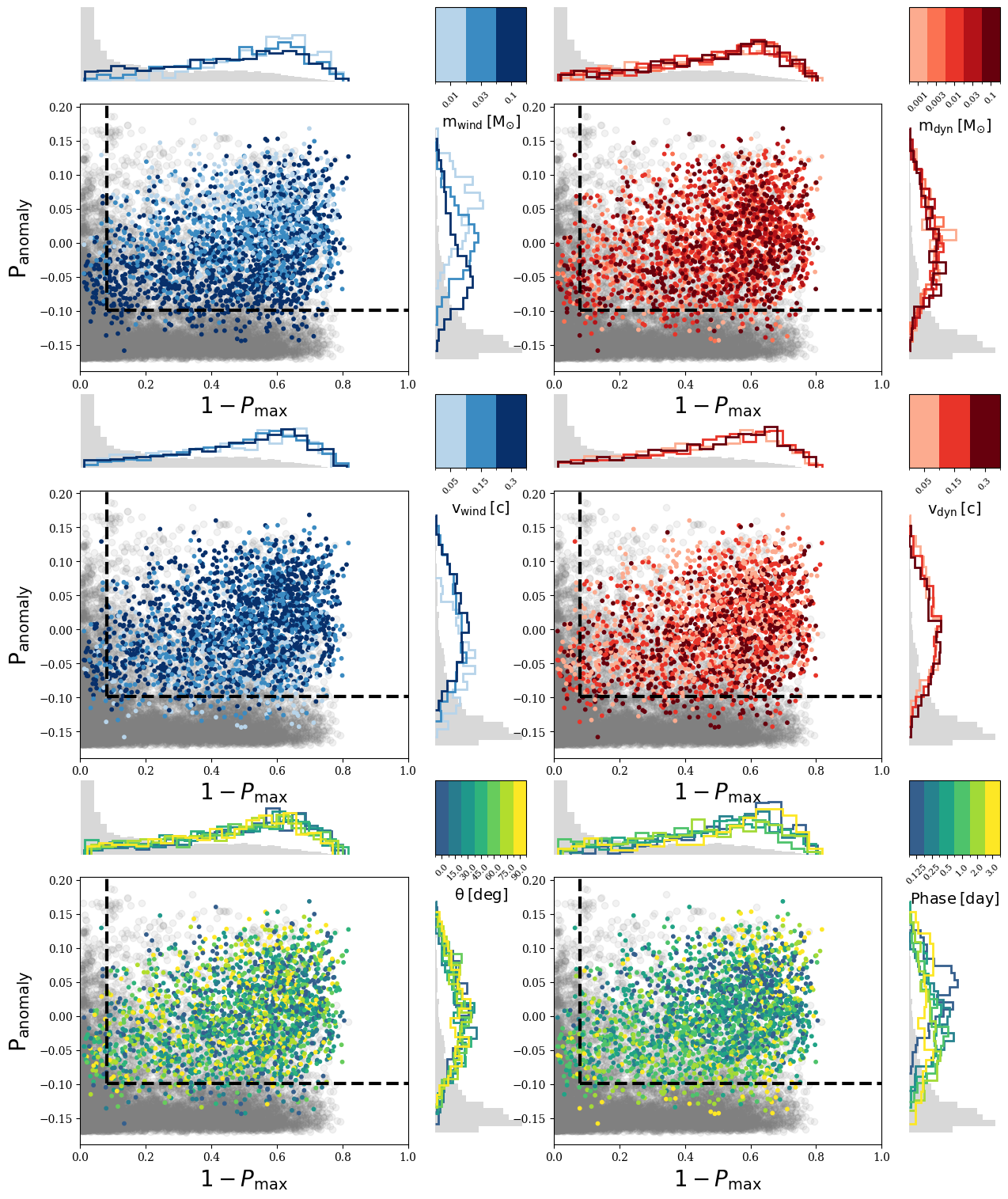}
    \caption{
    Hybrid framework decision space for the 20-filter configuration. 
    Non-KN test samples are shown in gray, while simulated KNe are overplotted and color-coded by ejecta parameters from the synthetic grid: wind ejecta mass ($m_{\rm wind}$), wind velocity ($v_{\rm wind}$), dynamical ejecta mass ($m_{\rm dyn}$), dynamical ejecta velocity ($v_{\rm dyn}$), viewing angle ($\theta$), and phase (days since merger). 
    The dashed lines indicate thresholds in $P_{ano}$ and $1-P_{\max}$ determined by maximizing Youden's~\,$J$ statistic. 
    Marginal histograms highlight the distributions of KNe in each parameter subset relative to the background of non-KN transients.
    }
    \label{fig:hybrid_kn_sim20}
\end{figure*}

\begin{figure*}
    \centering
    \includegraphics[width=0.8\linewidth]{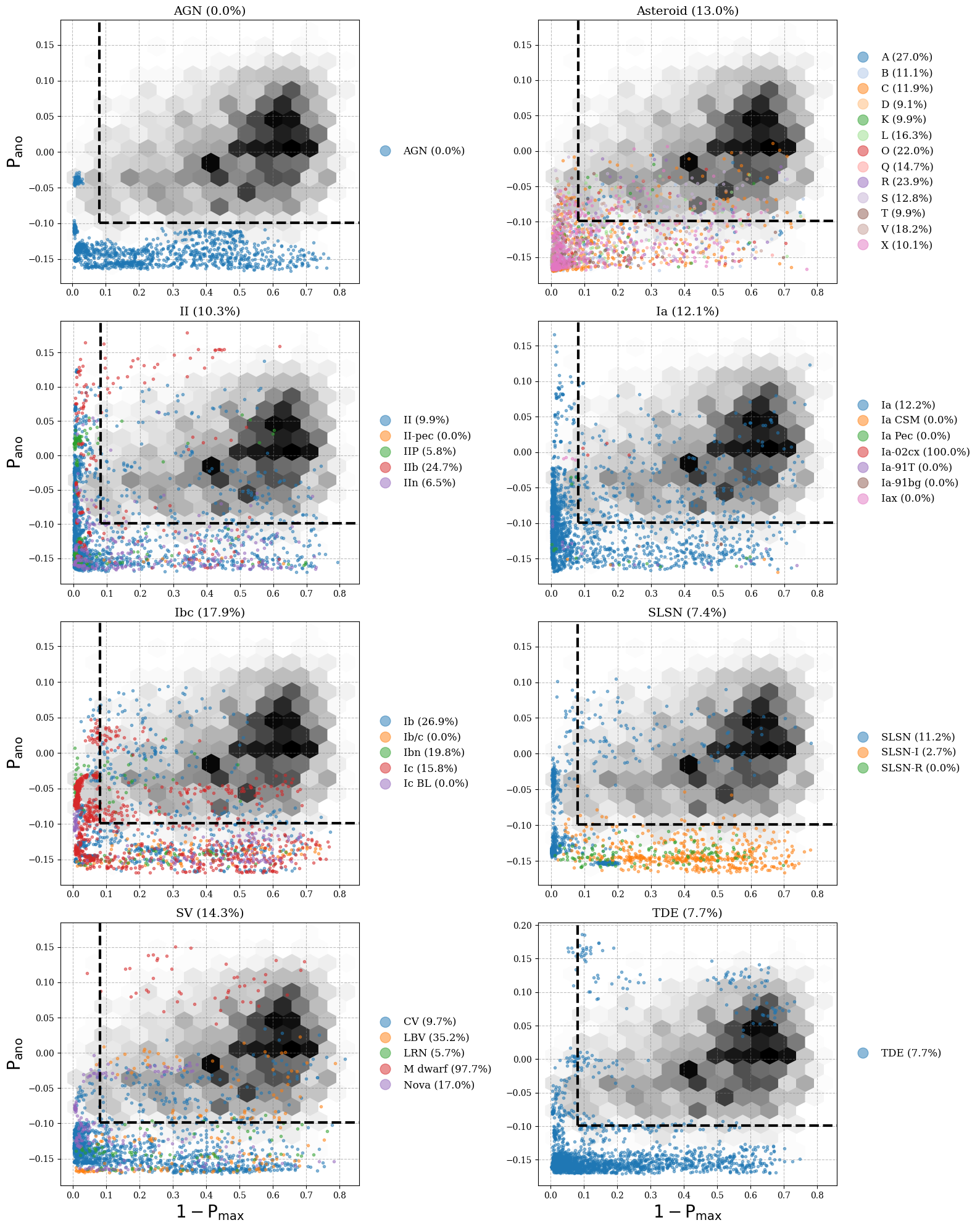}
    \caption{
    Distribution of non-KN test samples in the hybrid framework decision space for the 20 filter set configuration. 
    Each panel corresponds to one of the transient classes defined in Table~\ref{tab:type_mapping}, showing how the initially claimed subtypes from OSC and WISeREP are distributed with respect to the anomaly probability ($P_{\mathrm{ano}}$) and $1-P_{\max}$. 
    Colored points denote the subtypes redefined according to the simplified mapping used for this analysis, and the fraction in parentheses in each panel title indicates the percentage of sources classified as anomalies under the Youden's~\,$J$ threshold as described in Section~\,\ref{subsubsec:combined_decision}. 
    Legends on the right summarize the relative anomaly fractions for each subtype. 
    The gray hexagonal density background represents the distribution of simulated KNe, illustrating their characteristic region in the decision space. 
    }
    \label{fig:subtype_distribution_20}
\end{figure*}

\bibliography{sample7}{}
\bibliographystyle{aasjournalv7}



\end{document}